\documentclass[12pt]{article}
\usepackage{epsfig}
        \newdimen\eqskip
        \newdimen\txtskip
        \baselineskip=\txtskip
        \newdimen\mysep                
        \newdimen\hmysep
\begin{document}
       
  \newcommand{\ccaption}[2]{
    \begin{center}
    \parbox{0.85\textwidth}{
      \caption[#1]{\small{{#2}}}
      }
    \end{center}
    }
\newcommand{\BS}{\bigskip}
\def    \be             {\begin{equation}}
\def    \ee             {\end{equation}}
\def    \beq             {\begin{equation}}
\def    \eeq             {\end{equation}}
\def    \ba             {\begin{eqnarray}}
\def    \ea             {\end{eqnarray}}
\def    \beqn           {\begin{eqnarray}}
\def    \eeqn           {\end{eqnarray}}
\def    \nn             {\nonumber}
\def    \=              {\;=\;}
\def    \frac           #1#2{{#1 \over #2}}
\def    \ret            {\\[\eqskip]}
\def    \ie             {{\em i.e.\/} }
\def    \eg             {{\em e.g.\/} }
\def \lsim{\mathrel{\vcenter
     {\hbox{$<$}\nointerlineskip\hbox{$\sim$}}}}
\def \gsim{\mathrel{\vcenter
     {\hbox{$>$}\nointerlineskip\hbox{$\sim$}}}}
\def    \bentarrow      {\:\raisebox{1.1ex}{\rlap{$\vert$}}\!\rightarrow}
\def    \rd             {{\mathrm d}}    
\def    \Im             {{\mathrm{Im}}}  
\def    \bra#1          {\mbox{$\langle #1 |$}}
\def    \ket#1          {\mbox{$| #1 \rangle$}}

\def    \kev            {\mbox{$\mathrm{keV}$}}
\def    \mev            {\mbox{$\mathrm{MeV}$}}
\def    \gev            {\mbox{$\mathrm{GeV}$}}


\def    \mq             {\mbox{$m_Q$}}  
\def    \mt             {\mbox{$m_t$}}  
\def    \mb             {\mbox{$m_b$}}  
\def    \mqq            {\mbox{$m_{Q\bar Q}$}}
\def    \mqqsq          {\mbox{$m^2_{Q\bar Q}$}}
\def    \pt             {\mbox{$p_T$}}
\def    \ptsq           {\mbox{$p^2_T$}}

\newcommand     \MSB            {\ifmmode {\overline{\rm MS}} \else 
                                 $\overline{\rm MS}$  \fi}
\def    \muf            {\mbox{$\mu_{\rm F}$}}
\def    \mug            {\mbox{$\mu_\gamma$}}
\def    \mufsq          {\mbox{$\mu^2_{\rm F}$}}
\def    \mur            {{\mbox{$\mu_{\rm R}$}}}
\def    \mursq          {\mbox{$\mu^2_{\rm R}$}}
\def    \mul            {{\mu_\Lambda}}
\def    \mulsq          {\mbox{$\mu^2_\Lambda$}}

\def    \bzero          {\mbox{$b_0$}}
\def    \as             {\ifmmode \alpha_s \else $\alpha_s$ \fi}
\def    \asb            {\mbox{$\alpha_s^{(b)}$}}
\def    \assq           {\mbox{$\alpha_s^2$}}
\def \oacube {\mbox{$ O(\alpha_s^3)$}}
\def \oafour {\mbox{$ O(\alpha_s^4)$}}
\def \oatwo {\mbox{$ O(\alpha_s^2)$}}
\def \oas   {\mbox{$ O(\alpha_s)$}}
\def\slash#1{{#1\!\!\!/}}
\def\rt1{\raisebox{-1ex}{\rlap{$\; \rho \to 1 \;\;$}}
\raisebox{.4ex}{$\;\; \;\;\simeq \;\;\;\;$}}
\def\ltap{\raisebox{-.5ex}{\rlap{$\,\sim\,$}} \raisebox{.5ex}{$\,<\,$}}
\def\gtap{\raisebox{-.5ex}{\rlap{$\,\sim\,$}} \raisebox{.5ex}{$\,>\,$}} 
\begin{titlepage}
\nopagebreak
{\flushright{
        \begin{minipage}{5cm}
        CERN-TH/98-31\\
        {\tt hep-ph/9801375}\\
        \end{minipage}        }

}
\vfill
\begin{center}
{\LARGE { \bf \sc NLL Resummation \\[0.5cm]
of the Heavy-Quark \\[0.5cm] Hadroproduction Cross-Section}}
\vfill                                                       
{\bf Roberto BONCIANI, }
\\                                                                             
{INFN and Dipartimento di Fisica dell'Universit\`{a}, \\
 Largo E. Fermi 2, Firenze, Italy} 
\\[0.5cm] 
{\bf      Stefano CATANI
     \footnote{On leave of absence from INFN, Firenze, Italy},
          Michelangelo L. MANGANO                                
    \footnote{On leave of absence from INFN, Pisa, Italy}
     and Paolo NASON
    \footnote{On leave of absence from INFN, Milano, Italy} }\\
{CERN, Theoretical Physics Division, \\ CH~1211 Geneva 23, Switzerland} 
\end{center}                             
\nopagebreak
\vfill
\begin{abstract} 
We compute the effect of soft-gluon resummation,
at the next-to-leading-logarithmic level, in the
hadroproduction cross-section for heavy flavours.
Applications to top, bottom and charm total cross-sections
are discussed. We find in general that the corrections to the 
fixed next-to-leading-order results are
larger for larger renormalization scales, and small, or even negative, for
smaller scales. This leads to a significant reduction
of the scale-dependence
of the results, for most experimental configurations of interest.
\end{abstract}                                                
\vskip 1cm
CERN-TH/98-31\hfill \\
January 1998 \hfill  
\vfill       
\end{titlepage}
\section{Introduction}
\label{sec:1}\label{introduction}
Large  logarithmically-enhanced corrections due to soft-gluon radiation are a 
general feature in the study of the production cross-sections of high-mass 
systems near threshold. Techniques for resumming these corrections have been 
developed over the past several years, starting from the case of
Drell-Yan (DY) pair production~\cite{Sterman,CT}.             
In this paper we deal with the resummation of logarithmically-enhanced effects
in the vicinity of the threshold region in the total cross-section
for the hadroproduction of heavy quarks. This topic has been examined in
several studies in recent years~\cite{Laenen}--\cite{CMNT2}. These
analyses were limited to soft-gluon resummation to leading-logarithmic 
accuracy. The work we present here extends and updates the results of  
refs.~\cite{cmnt21} and \cite{CMNT2} by consistently including all the
next-to-leading logarithmic contributions.

The resummation program of soft-gluon contributions is best carried 
out in the Mellin-transform space, or $N$-space. 
In $N$-moment space the threshold-production region corresponds to the limit
$N\to \infty$ and the typical structure of the threshold corrections
is as follows                                            
\beq
\label{genlog}
{\hat \sigma}_N^{(LO)} \left\{ 1 + \sum_{n=1}^{\infty} \as^n \sum_{m=1}^{2n}
c_{n,m} \ln^m N \right\} \;\;,
\eeq 
where ${\hat \sigma}_N^{(LO)}$ is the corresponding partonic cross-section at
leading order (LO). In the case of hadronic collisions, the best studied
process~\cite{Sterman,CT} is Drell-Yan (DY) lepton-pair production. In the DY
process the logarithmic terms in the curly bracket of eq.~(\ref{genlog})
can be explicitly summed and organized in a radiative factor $\Delta_{DY,N}$
that has an exponential form:
\beqn 
\label{deltady}
\Delta_{DY,N}(\as) &=& \exp \left\{ \sum_{n=1}^{\infty} \as^n 
\sum_{m=1}^{n+1} G_{nm} \ln^m N \right\} \\
\label{deltadyex}
&=& \exp \left\{ \ln N \,g_{DY}^{(1)}(\as \ln N) +  g_{DY}^{(2)}(\as \ln N)
+ \as g_{DY}^{(3)}(\as \ln N) + \dots \right\} \;\;.
\eeqn
Note that the exponentiation in eq.~(\ref{deltady}) is not trivial. The sum
over $m$ in eq.~(\ref{genlog}) extends up to $m=2n$ while in
eq.~(\ref{deltady}) the maximum value for $m$ is smaller, $m \leq n+1$. In
particular, this means that all the double logarithmic (DL) terms
$\as^n c_{n,2n} \ln^{2n} N$ in eq.~(\ref{genlog}) are taken into account
by simply exponentiating the lowest-order contribution 
$\as c_{1,2} \ln^2 N$. Then, the exponentiation in eq.~(\ref{deltady})
allows one to define the improved perturbative expansion in 
eq.~(\ref{deltadyex}). The function $\ln N \,g_{DY}^{(1)}$
resums all the {\em leading} logarithmic (LL) contributions 
$\as^n \ln^{n+1} N$, $g_{DY}^{(2)}$ contains the {\em next-to-leading}
logarithmic (NLL) terms $\as^n \ln^n N$,
$\as g_{DY}^{(3)}$ contains the {\em next-to-next-to-leading}
logarithmic (NNLL) terms $\as^{n+1} \ln^n N$,
and so forth. Once the functions
$g_{DY}^{(k)}$ have been computed, one has a systematic perturbative
treatment of the region of $N$ in which $\as \ln N \ltap 1$, which is much
larger than the domain $\as \ln^2 N \ll 1$ in which the fixed-order calculation
in $\as$ is reliable.
 
The physical bases for the soft-gluon exponentiation in the DY process
are dynamics and kinematics factorizations~\cite{Sterman,CT}. 
Dynamics factorization follows
from gauge invariance and unitarity: in the soft limit multi-gluon amplitudes
fulfil generalized factorization formulae given in terms of a single-gluon
emission probability. Kinematics factorization follows from the fact that
the multi-gluon phase space in the appropriate soft limit can  
exactly be factorized by working in $N$ space. 

The general extension of these results to the resummation of NLL contributions
in heavy-quark production was first 
considered by Kidonakis and Sterman~\cite{kidon} and more recently examined
in ref.~\cite{bonciani}. There are kinematics and dynamics
complications in going from the DY to the heavy-quark processes. Although in
the heavy-quark case  the 
kinematics is more involved 
it can nonetheless be recast in factorized form in $N$-moment 
space~\cite{kidon,bonciani}. 
Dynamics complications are more essential. While in the DY case
resummation is only associated with the $q\bar{q}$-annihilation subprocess,
in the heavy-quark case
one has to consider both the $q\bar{q}$ and $gg$ subprocesses. Moreover,
unlike the DY vector boson, the heavy-quark pair carries colour charge. Thus
the heavy-quark cross-section is sensitive to radiative corrections due to 
soft-gluon emission from the heavy quark and antiquark produced in the
final state. In particular, this emission induces colour exchange in the hard
subprocess, whose colour state can thus fluctuate. The presence of 
colour fluctuations implies that, after soft-gluon resummation, the 
partonic cross-sections have an exponentiated form that is more involved
than the simple exponential in eq.~(\ref{deltadyex}).

The general structure of the resummed partonic cross-section for heavy-quark
hadroproduction is as follows~\cite{kidon}
\beq                                      
\label{genresHQ}
\hat{\sigma}_{ij, N}^{(res)} =
\sum_{{\bf I},{\bf J}} M_{ij,{\bf I}, \,N}^\dagger \;
\left[ {\bf \Delta}_{ij, N} \right]_{{\bf I}, {\bf J}}
\; M_{ij,{\bf J}, \,N} \;\;,
\eeq
where all the various factors depend on the heavy-quark mass $m$ and the
renormalized coupling $\as(\mu^2)$ evaluated at the scale $\mu$, although 
this dependence is not explicitly denoted. The sum in eq.~(\ref{genresHQ})
is over all possible colour states ${\bf I},{\bf J}$ of the LO hard-scattering
subprocess $i + j \to Q {\overline Q}$. The colour amplitude 
$M_{ij,{\bf I}, \,N}$
is computable as a power series expansion in $\as$, but it does not contain
any logarithmically-enhanced corrections produced by soft-gluon radiation.
These corrections are embodied by the factor ${\bf \Delta}_{ij, N}$ that is
a matrix in the space of the colour states. Performing the
all-order resummation,
the factor ${\bf \Delta}_{ij, N}$ can be expressed in exponential form but,
unlike $\Delta_{DY, N}$, we are now dealing with an exponential matrix and
exponentiation has to be understood in a formal sense. The analogues of
the functions $g_{DY}^{(k)}$ are colour matrices. They have to be evaluated
and then diagonalized, before one can rewrite the right-hand side of 
eq.~(\ref{genresHQ}) as an actual exponential.
 
The resummation formula (\ref{genresHQ}) applies to the general case of
heavy-quark production at fixed invariant mass of the produced quark pair and
at fixed scattering angle~\footnote{A similar structure also holds for the
production of dijets at fixed invariant mass~\cite{jetnlo}.}.
In this case, the radiative factor ${\bf \Delta}_{ij, N}$ depends
on the difference in rapidity between the heavy quark and antiquark
and the NLL resummation has been worked out explicitly in ref.~\cite{kidon}.

The goal of this paper is the resummation of 
logarithmically-enhanced corrections to the heavy-quark {\em total}
hadroproduction cross-section, up to NLL order.
The non-trivial angular dependence of the eigenvalues and eigenvectors    of
the radiative factor ${\bf \Delta}_{ij, N}$ calculated in ref.~\cite{kidon}, 
however, makes it difficult to 
apply in practice~\cite{KSV} 
this approach to the evaluation of the total production
cross-section. 
A new general method for the resummation of 
soft-gluon contributions to hard-scattering processes has recently been 
developed in refs.~\cite{bonciani,inprep}. This technique allows a more direct
evaluation of the logarithmically-enhanced corrections to the heavy-quark {\em
total} hadroproduction cross-section, up to NLL accuracy\footnote{We have
checked that our resummation    formulae are consistent with the NLL results on
the heavy-quark cross-section at fixed invariant mass obtained in
ref.~\cite{kidon}.}.  
A key simplification in this method is the direct
diagonalization of the radiative           
factor ${\bf \Delta}_{ij, N}$ for the resummed parton-level total cross-section
in the colour-basis defined by colour-singlet and colour-octet final 
states. Details on                                
the calculation and the illustration of the general method are presented 
elsewhere~\cite{bonciani,inprep}.                              
In this paper we limit ourselves to presenting the final theoretical 
results on soft-gluon resummation to NLL order, and we mainly concentrate on
the ensuing phenomenological predictions. 
                        
The rest of this paper is organized as follows. In Section~\ref{notation}
we describe our notation, and collect the fixed-order leading
and next-to-leading formulae that will be used in the following.
In Section~\ref{resumsec} we give the formulae for the NLL resummed
cross-sections, and in Sect. \ref{Results} we present our phenomenological
results.

\section{Notation and fixed-order calculations}
\label{sec:2}\label{notation}
In this Section we recall the known theoretical results~\cite{NDE,VN}
on the calculation of the heavy-quark hadroproduction cross-section
at LO and next-to-leading-order (NLO) in QCD perturbation theory.
We shall follow closely the notation of ref.~\cite{NDE}.

The total cross-section $\sigma$ is given by the following factorization
formula
\beq \label{HVQCrossSection}
\sigma(\rho_h,m^2) = \sum_{i,j} \int_0^1 dx_1\,dx_2 
\;F_i(x_1,\mu^2) \;F_j(x_2,\mu^2)
\;{\hat \sigma}_{ij}
\left(\rho;m^2,\as(\mu^2),\mu^2\right)\;,
\eeq
where $m$ is the mass of the heavy quark, $S$ is the square of the
centre-of-mass energy, $i,j$ denote parton indices $(i=q,{\bar q},g)$ and
the dimensionless variables $\rho$ and $\rho_h$ are
\beq
\rho_h=\frac{4m^2}{S}\,, \quad \rho=\frac{\rho_h}{x_1\,x_2}\,.
\eeq
The parton densities $F_i(x,\mu^2)$ and the partonic cross-sections
${\hat \sigma}_{ij}$ depend on the factorization scale $\mu$ (which we always
set equal to the renormalization scale) and on the factorization scheme.
We use the \MSB\ factorization scheme.

We also introduce the dimensionless functions $f_{ij}$ by rescaling
the partonic cross-sections as follows
\beq\label{sighq}
{\hat \sigma}_{ij}(\rho;m^2,\as(\mu^2),\mu^2) \equiv 
\frac{\as^2(\mu^2)}{m^2} \;
f_{ij}(\rho;\as(\mu^2),\mu^2/m^2) \;\;.
\eeq
The functions $f_{ij}$ are perturbatively computable and given by the
following expansion in $\as$
\beq\label{fpert}                
f_{ij}(\rho;\as(\mu^2),\mu^2/m^2) = f_{ij}^{(0)}(\rho) + 
4\pi \as(\mu^2) \left[
f_{ij}^{(1)}(\rho) + {\overline f}_{ij}^{(1)}(\rho)
\ln \frac{\mu^2}{m^2} \right] 
+ \sum_{n=2}^{\infty} \as^n(\mu^2) f_{ij}^{(n)}(\rho;\mu^2/m^2) \;\;.
\eeq
The 
LO terms $f_{ij}^{(0)}$ in eq.~(\ref{fpert}) are 
explicitly given by
\beqn\label{fqqbar}
f_{q{\bar q}}^{(0)}(\rho) &=& \frac{\pi}{6} \;\frac{T_R C_F}{N_c} \;
\beta \rho \;( 2 + \rho) \;\;,
\\
\label{fgg}
f_{gg}^{(0)}(\rho) &=& \frac{\pi}{12} \;\frac{T_R}{N_c^2 -1} \;
\beta \rho  \left\{ 3 C_F \left[ ( 4 + 4 \rho - 2 \rho^2) \;
\frac{1}{\beta} \;
\ln \frac{1+\beta}{1-\beta} - 4 - 4 \rho \right] \right. \nonumber \\
&+& C_A \left. \left[ 3 \rho^2 \;\frac{1}{\beta} \;\ln \frac{1+\beta}{1-\beta}
- 4 - 5 \rho \right] \right\} \;\;,
\eeqn
where $\beta \equiv \sqrt {1-\rho} \;$
and $f_{ij}^{(0)}(\rho) = 0$ for all the other parton channels.

The NLO terms ${\overline f}_{ij}^{(1)}(\rho)$ in eq.~(\ref{fpert})
are obtained by simply performing the convolution of $f_{ij}^{(0)}(\rho)$
with Altarelli-Parisi splitting functions. The remaining NLO contributions
$f_{ij}^{(1)}(\rho)$, first evaluated in ref.~\cite{NDE}, are not known in
analytic form but only numerically~\cite{NDE,VN}. An analytic parametrization
of the numerical results is available in ref.~\cite{NDE}. The functions
$f_{q{\bar q}}^{(1)}(\rho)$ and $f_{gg}^{(1)}(\rho)$ are plotted in 
fig.~\ref{fig:qqnlo} (solid lines).

\subsection{Threshold behaviour}                                      
We are mainly interested in the behaviour of QCD corrections near the
threshold region, $\rho \to 1$. In this region, the LO functions
$f_{ij}^{(0)}(\rho)$ vanish because of phase-space suppression:
\beq
f^{(0)}_{q {\bar q}}(\rho) \rt1 \frac{T_R C_F}{2N_c} 
\,  \pi \beta \to 0 \;\;,
\quad
f^{(0)}_{gg}(\rho) \rt1 \frac{T_R}{N_c^2-1} \, 
\left( C_F- \frac{C_A}{4} \right) \pi \, \beta \to 0 \;\;,
\eeq
while the NLO functions 
do not. The behaviour of the latter is analytically known and given by 
the following expressions~\cite{NDE,VN}
\beqn
f_{q {\bar q}}^{(1)}(\rho) + {\overline f}_{q {\bar q}}^{(1)}(\rho)
\ln \frac{\mu^2}{m^2} &\!=\!& \frac{1}{4\pi^2} f^{(0)}_{q \bar{q}}(\rho) 
\left\{ \left( C_F- \frac{1}{2} C_A \right)
\, \frac{\pi^2}{2\beta} + 2C_F \, \ln^2 (8 \beta^2) \right. \label{f1qq} \\
&\!-\!& \!\! \left. (8C_F+C_A) \, \ln (8 \beta^2)  - 2C_F \, \ln (4 \beta^2) 
\, \ln \frac{\mu^2}{m^2}+{\overline C}_2\!\left(\frac{\mu^2}{m^2}\right)
+{\cal O}(1- \rho) \right\} \, , \nonumber \\
f_{gg}^{(1)}(\rho) + {\overline f}_{gg}^{(1)}(\rho)
\ln \frac{\mu^2}{m^2} &\!=\!& \frac{1}{4\pi^2} f^{(0)}_{gg}(\rho)
\left\{ \frac{N_c^2+2}{N_c (N_c^2-2)} \frac{\pi^2}{4 \beta} +2 C_A \, 
\ln^2 (8 \beta^2) \right. \label{f1gg} \\
&\!-\!& \!\! \left. \frac{(9N_c^2-20)C_A}{N_c^2-2} \, \ln (8 \beta^2)
-2 C_A \, \ln (4 \beta^2) \, \ln \frac{\mu^2}{m^2} 
+{\overline C}_3\!\left(\frac{\mu^2}{m^2}\right) +{\cal O}(1- \rho) 
\right\}
\, , \nonumber
\eeqn
and $f_{ij}^{(1)}(\rho), {\overline f}_{ij}^{(1)}(\rho) = {\cal O}(\beta)$
for all the other parton channels.

The first term in the curly brackets of eqs.~(\ref{f1qq},\ref{f1gg}) is due
to the Coulomb interaction between the heavy quark and antiquark. The remaining
logarithmic contributions are produced by soft-gluon bremsstrahlung. Note
that the right-hand sides of eqs.~(\ref{f1qq},\ref{f1gg}) also include the
constant coefficients ${\overline C}_2, {\overline C}_3$, which are due to
large-momentum virtual corrections. The explicit expressions for these
coefficients are
\beq
\label{cbarcoef}
{\overline C}_2\!\left(\frac{\mu^2}{m^2}\right) 
= C_2 + (n_{lf}-4) \left( \frac{2}{3} \ln 2
- \frac{5}{9} \right)
+ \left( \frac{53}{6} - \frac{1}{3} n_{lf} \right) \ln{\frac{\mu^2}{m^2}} 
\,, \quad
{\overline C}_3\!\left(\frac{\mu^2}{m^2}\right) = 
C_3 +12 \, \ln{\frac{\mu^2}{m^2}} \, ,
\eeq
where $n_{lf}$ denotes the number of massless flavours (i.e. the number of
quark flavours that are lighter than the produced heavy-quark pair)  
and $C_{2,3}$ are obtained from the values of the constants
$a_0$ reported in Table~1 of ref.~\cite{NDE}: 
\be
\label{ndecoef}            
   C_2=36 \pi a_0^{q\bar q} \simeq 20.45 \;\;, \quad \quad \quad 
   C_3=\frac{768}{7} \pi a_0^{gg} \simeq 37.23 \;\;.
\ee 

In the strict threshold limit the Coulomb contributions dominate the radiative
corrections in eqs.~(\ref{f1qq},\ref{f1gg}). However, in most of the
hadroproduction processes of phenomenological interest, the heavy-quark pair 
is not produced exactly at threshold. In these cases, it turns out~\cite{NDE,VN}
that the logarithmic terms due to soft-gluon emission 
give the bulk of the NLO corrections. Multiple-gluon radiation 
at higher perturbative orders leads to more enhanced logarithmic corrections
and the coefficient functions $f_{ij}^{(n)}(\rho;\mu^2/m^2)$ in
eq.~(\ref{fpert}) behave as
\beq
f_{ij}^{(n)}(\rho;\mu^2/m^2) \sim f_{ij}^{(0)}(\rho) \ln^{2n} \beta^2 \;\;.
\eeq
Resummation of these soft-gluon effects to all orders in perturbation theory
can be important to improve the reliability of the QCD predictions.
 
\subsection{$N$-moment space}
\label{Nsec}
The resummation program of soft-gluon contributions has been carried 
out~\cite{Sterman,CT} in the Mellin-transform space, or $N$-space. 
Working in $N$-space, one can disentangle the soft-gluon effects in the parton
densities from those in the partonic cross-section and one can
straightforwardly implement and factorize the kinematic constraints of
energy and longitudinal-momentum conservation. 

The Mellin transform of the heavy-quark hadronic cross-section
$\sigma(\rho_h,m^2)$ is defined as follows
\beq
\sigma_N(m^2) \equiv \int_0^1 d\rho \;\rho^{N-1} \;\sigma(\rho,m^2) \;\;,
\eeq
where $N$ is the moment index. In $N$-moment space eq.~(\ref{HVQCrossSection})
takes a simple factorized expression
\beqn \label{HVQCSN}
\sigma_N(m^2) &=& \sum_{i,j}  \;F_{i,N+1}(\mu^2) \;F_{j,N+1}(\mu^2)
\;{\hat \sigma}_{ij,N}(m^2,\as(\mu^2),\mu^2) \\
\label{HVQfN}
&=& \sum_{i,j}  \;F_{i,N+1}(\mu^2) \;F_{j,N+1}(\mu^2)  
\;\frac{\as^2(\mu^2)}{m^2} \;f_{ij,N}(\as(\mu^2),\mu^2/m^2) \;\;,
\eeqn
where we have used eq.~(\ref{fpert}) and we have 
introduced the $N$-moments $F_{i,N}(\mu^2)$ of the parton
densities,
\beq 
F_{i,N}(\mu^2) \equiv \int_0^1 dx \;x^{N-1} \;F_i(x,\mu^2) \;\;,
\eeq
and likewise for any other $x$-space function.

The $N$-moments of the LO functions in eqs.~(\ref{fqqbar}) and (\ref{fgg})
are given by the following explicit expressions~\cite{CMNT2}
\beqn\label{fqqbarN}
f_{q{\bar q}, \,N}^{(0)} &=& \frac{\pi^{\frac{3}{2}}}{4} \;
\frac{T_R C_F}{N_c} \; \frac{\Gamma(N+1)}{\Gamma(N + 7/2)} \; (N+2) \;\;,
\\ \label{fggN}
f_{gg, \,N}^{(0)} &=& \frac{\pi^{\frac{3}{2}}}{4} \;
\frac{T_R}{N_c^2 -1} \;\frac{\Gamma(N+1)}{\Gamma(N + 5/2)} \;\frac{1}{N+3}
\nonumber \\
&\times& \left[ 2 C_F \;\frac{N^3 + 9 N^2 + 20 N + 14}{(N+1)(N+2)}
\; - C_A \;\frac{N^2 + 8 N + 11}{2 N + 5} \right] \;\;.
\eeqn

To discuss the structure of the higher-order corrections considered 
in the next Section, it is useful to decompose the various 
cross-section contributions according to the colour state
(singlet ${\bf 1}$ and octet ${\bf 8}$)
of the heavy-quark pair produced in the final state.
When applied to the LO functions $f_{ij}^{(0)}$, this decomposition gives
\beq \label{fij018}
f_{ij}^{(0)}(\rho) = f_{ij,{\bf 1}}^{(0)}(\rho) + f_{ij,{\bf 8}}^{(0)}(\rho)
\eeq
with~\cite{cmnt21}
\beqn
&&f_{gg,{\bf 1}}^{(0)}(\rho)=  \frac{T_R}{N_c(N_c^2-1)}
\frac{\pi \beta \rho}{8}  \left[ ( 4 + 4 \rho - 2 \rho^2) \;
\frac{1}{\beta} \;\ln \frac{1+\beta}{1-\beta} - 4 - 4 \rho \right] ,
\nonumber \\ \label{f0colour}
&&f_{gg,{\bf 8}}^{(0)}(\rho)= f_{gg}^{(0)}(\rho) -
f_{gg,{\bf 1}}^{(0)}(\rho) \;,
\quad f_{q{\bar q},{\bf 1}}^{(0)}(\rho) = 0 \;,
\quad f_{q{\bar q},{\bf 8}}^{(0)}(\rho) = f_{q{\bar q}}^{(0)}(\rho) \;.
\eeqn
The $N$-moments of $f_{gg,{\bf 1}}^{(0)}(\rho)$ are
\beq
\label{fgg01}
f_{gg,{\bf 1}, \,N}^{(0)} =  \frac{\pi^{\frac{3}{2}}}{4} \;
\frac{T_R}{N_c(N_c^2 -1)} \;\frac{\Gamma(N+1)}{\Gamma(N + 5/2)} 
\;\frac{N^3 + 9 N^2 + 20 N + 14}{(N+1)(N+2)(N+3)} \;\;,
\eeq
and those of $f_{gg,{\bf 8}}^{(0)}(\rho)$ and 
$f_{q{\bar q},{\bf 8}}^{(0)}(\rho)$ are then obtained from 
eqs.~(\ref{fqqbarN},\ref{fggN}) by using 
eq.~(\ref{f0colour})

We also introduce the functions $f_{ij}^{(1),{\rm Coul}}(\rho)$ that describe
the Coulomb contributions to the NLO corrections in                 
eq.~(\ref{f1qq},\ref{f1gg}). Using the singlet/octet colour decomposition,
we have
\beqn\label{f1singC}
f_{ij,{\bf 1}}^{(1),{\rm Coul}}(\rho) &=& 
f_{ij,{\bf 1}}^{(0)}(\rho) \; C_F \; \frac{1}{8\beta} \;\;,
\\
\label{f1octC}
f_{ij,{\bf 8}}^{(1),{\rm Coul}}(\rho) &=& 
f_{ij,{\bf 8}}^{(0)}(\rho) \; \left(C_F-\frac{C_A}{2}\right) 
\; \frac{1}{8\beta} 
\;\;,
\eeqn 
whose Mellin transforms are given by  
\beqn\label{f1qqC}                    
f_{q\bar q,{\bf 8}, \, N}^{(1),{\rm Coul}} &=& - \frac{\pi}{96} 
\frac{T_R C_F}{N_c^2} \frac{3N+5}{(N+1)(N+2)} \; ,
\\
\label{f1ggOC}
f_{gg,{\bf 8}, \, N}^{(1),{\rm Coul}} &=& - \frac{\pi}{192} \frac{T_R}{N_c (N_c^2-1)}
\left[ N_c \left( 3 I_{N+2} -\frac{4}{N+1} - \frac{5}{N+2} \right) \right.
\nn \\
&+& \left. 3 \;\frac{N_c^2-2}{N_c} \left( 2I_N +  2I_{N+1} - I_{N+2}
- \frac{2}{N+1}-\frac{2}{N+2} \right) \right] \;\;,\\
\label{f1ggSC}
f_{gg,{\bf 1}, \, N}^{(1),{\rm Coul}} &=& \frac{\pi}{64} \frac{T_R}{N_c^2}
\left[  2I_N +  2I_{N+1} - I_{N+2} - 
\frac{2}{N+1}-\frac{2}{N+2}\right] \; .
\eeqn                                              
The function $I_N$ is defined by:
\be
I_N \; = \; \int_0^1 d\rho \;\rho^N \frac{1}{\sqrt {1-\rho}}
\ln \frac{1 + \sqrt {1-\rho}}{1 - \sqrt {1-\rho}} 
\label{INdef1}
\ee          
We can rewrite $I_N$ as follows
\be                         
I_N \; = \; \int_0^1 d\rho \;\rho^N \frac{1}{\sqrt {1-\rho}}
\left[ \ln \frac{1}{\rho} \, + \, 
       2\ln (1 + \sqrt{1-\rho}) \right] \; .                 
\ee                                         
The integral of the first term can be done exactly, while to integrate 
the second term we first expand the logarithm of $1+\sqrt{1-\rho}$ as:
\be\label{logoneplusx}                                           
      \ln(1+x) \= \sum_{n=1}^{4} \alpha_n x^n \; ,
\ee
with $\alpha_n=0.9991,-0.4828,0.2477,-0.0712$ for $n=1,\dots,4$,
and then perform the integral term by term.                     
The expansion in eq.~(\ref{logoneplusx})
is accurate to $10^{-4}$ in the needed range $0<x<1$.     
The final approximate result for $I_N$ is:
\be
I_N \;\simeq\; \sqrt{\pi} \frac{\Gamma(N+1)}{\Gamma(N+3/2)} 
       \left[ \psi(N+3/2) \; - \; \psi(N+1) \right] 
      + 2 \sum_{n=1}^{4} \alpha_n
\frac{\Gamma(N+1)\Gamma(\frac{n+1}{2})}{\Gamma(N+n/2+3/2)} \; .
\label{INdef2}                                                 
\ee    

The threshold region $\rho \to 1$ corresponds to the limit $N \to \infty$
in $N$-moment space. In this limit the Mellin transform of 
the threshold expansions (\ref{f1qq},\ref{f1gg}) of 
the NLO corrections can be easily computed.
Having introduced the Coulomb contributions $f_{ij}^{(1),{\rm Coul}}$, we 
can write the result as follows
\beqn     
f_{q {\bar q}, N}^{(1)} + {\overline f}_{q {\bar q}, N}^{(1)}
\ln \frac{\mu^2}{m^2} &=& 
\frac{1}{4\pi^2} f^{(0)}_{q \bar{q},N} \left\{  
2C_F \ln^2 N + \left( C_A + 4C_F \gamma_E +2C_F \ln \frac{\mu^2}{4m^2}
\right) \ln N  \right. \nn \\
&+& \left. C_{q{\bar q}}(\mu^2/m^2) +
{\cal O}\left(\frac{1}{N} \right) \right\} 
+ f_{q\bar q,{\bf 8}, \, N}^{(1),{\rm Coul}} \;\;, 
\label{f1qqN}\\
f_{gg, N}^{(1)} + {\overline f}_{gg, N}^{(1)}
\ln \frac{\mu^2}{m^2} &=& 
\frac{1}{4\pi^2} f^{(0)}_{gg, N} \left\{  
2C_A \ln^2 N + \left[ \frac{N_c^2-4}{N_c^2-2} C_A + 4C_A \gamma_E
+ 2C_A \ln \frac{\mu^2}{4m^2} \right] \, \ln N  \right. \nn \\
&+& \left. C_{gg}(\mu^2/m^2) + {\cal O}\left(\frac{1}{N} \right) \right\}
+  f_{gg,{\bf 1}, \, N}^{(1),{\rm Coul}} + f_{gg,{\bf 8}, \, N}^{(1),{\rm Coul}} 
\label{f1ggN}\;\;,
\eeqn
where $\gamma_E=0.5772\ldots$ is the Euler number and the constant coefficients
$C_{q{\bar q}}, C_{gg}$ are related to those in       
eqs.~(\ref{cbarcoef},\ref{ndecoef}):
\ba  
 C_{q\bar q}(\frac{\mu^2}{m^2})
 &\!\!=\!\!\!& {\overline C}_2(\frac{\mu^2}{m^2}) + 
  2 C_F \left[\ln^2 2 +(\gamma_E-2)\ln \frac{\mu^2}{4m^2}+\frac{\pi^2}{2}
 +\gamma_E(\gamma_E-4)\right] \nn \\
 &+& (8C_F+C_A) (\gamma_E-2-\ln{2}) \;\;, \label{coefqq}\\
 C_{gg}(\frac{\mu^2}{m^2}) &\!\!=\!\!\!& {\overline C}_3(\frac{\mu^2}{m^2}) + 
  2 C_A \left[ \ln^2 2 +(\gamma_E-2)\ln \frac{\mu^2}{4m^2}+\frac{\pi^2}{2}  
 +\gamma_E (\gamma_E-4) \right] \nn \\
 &+& \frac{C_A(9N_c^2-20)}{N_c^2-2}
 (\gamma_E-2-\ln 2) \;\;. \label{coefgg}
\ea                       

The double and single logarithmic terms $\ln^2N$ and $\ln N$ in 
eqs.~(\ref{f1qqN},\ref{f1ggN}) are produced by soft-gluon radiation. These 
terms dominate the corrections in the curly brackets of 
eqs.~(\ref{f1qqN},\ref{f1ggN}). 
The resummation of soft-gluon contributions to all orders in
perturbation theory is considered in the following Section. 

\section{Soft-gluon resummation}
\label{resumsec}
The partonic cross-sections ${\hat \sigma}_{ij, N}$ with $ij \neq q{\bar q},
gg$ vanish at LO and are hence suppressed by a factor of $\as$ with respect to
${\hat \sigma}_{q{\bar q}, N}$ and ${\hat \sigma}_{gg, N}$. In the threshold
or large-$N$ limit, this relative suppression is further enhanced by a factor
of ${\cal O}(1/N)$. Therefore, we make no attempt to resum soft-gluon
corrections to these partonic channels and we restrict ourselves to
${\hat \sigma}_{q{\bar q}, N}$ and ${\hat \sigma}_{gg, N}$.
\subsection{Total heavy-quark cross-section to NLL accuracy: the radiative
factors}
\label{radfact}
As discussed in the Introduction, the resummed radiative factor 
$\left[ {\bf \Delta}_{ij, N} \right]_{{\bf I}, {\bf J}}$
in eq.~(\ref{genresHQ}) depends in general on the partonic subprocess
(labelled by $ij$) and on its colour states (labelled by $\bf I,J$). 
However, in the case of the threshold behaviour of the {\em total} cross 
section, most of the complications related to the colour structure can 
be easily overcome. Up to NLL accuracy, it turns out~\cite{bonciani,inprep}
that the colour
matrix ${\bf \Delta}_{ij, N}$ is {\em diagonal} with respect to the basis
$\bf I = \{ 1 , 8 \}$, $\bf 1$ and $\bf 8$ being the singlet and octet 
state of the produced heavy-quark pair. We can thus present the exponentiated
formulae for its eigenvalues $\Delta_{ij,{\bf I}, \, N}$
in the two colour channels.

We use a notation similar to that in ref.~\cite{CT} to facilitate the 
comparison with the known results for the DY process. In the
$\MSB$ factorization and renormalization schemes,
the NLL resummed expressions for the radiative factors 
$\Delta_{ij,{\bf I}, \, N}$ are:
\ba
\ln \Delta_{q \bar{q},{\bf 1}, \,N} &=& \int_{0}^{1} dz \, \frac{z^{N-1}-1}{1-z}
\,  \int_{\mu^2}^{4m^2(1-z)^2}
\frac{dq_{\perp}^2}{q_{\perp}^2}
\, 2A_q \left( \as(q_{\perp}^2) \right) + {\cal O}(\as (\as \ln N)^k) \;, 
\label{dqq1}\\
\ln \Delta_{q \bar{q},{\bf 8},N} &=& \int_{0}^{1} dz \, \frac{z^{N-1}-1}{1-z} 
\, \left\{ \int_{\mu^2}^{4m^2(1-z)^2}
\frac{dq_{\perp}^2}{q_{\perp}^2} 
\, 2A_{q} \left( \as(q_{\perp}^2) \right) 
+ D_{Q \bar{Q}} \left( \as \left( 4m^2(1-z)^2 \right) \right) \right\} 
\nn \\
&+& {\cal O}(\as (\as \ln N)^k) \;, \label{dqq8} \\
\ln \Delta_{gg,{\bf 1}, \,N} &=& \int_{0}^{1} dz \, \frac{z^{N-1}-1}{1-z} \, 
\int_{\mu^2}^{4m^2(1-z)^2}
\frac{dq_{\perp}^2}{q_{\perp}^2} \, 2A_g \left( \as(q_{\perp}^2) \right) 
+ {\cal O}(\as (\as \ln N)^k) \;,
\label{dgg1} \\
\ln \Delta_{gg,{\bf 8}, \,N} &=& \int_{0}^{1} dz \, \frac{z^{N-1}-1}{1-z} \, 
\left\{ \int_{\mu^2}^{4m^2(1-z)^2}
\frac{dq_{\perp}^2}{q_{\perp}^2} \, 2A_g \left( \as(q_{\perp}^2) \right) 
+ D_{Q \bar{Q}} \left( \as \left( 4m^2(1-z)^2 \right) \right) \right\}
\nn \\
&+& {\cal O}(\as (\as \ln N)^k) \;, \label{dgg8} 
\ea
where $A_q$~\cite{KT}, $A_g$~\cite{CdET}, $D_{Q \bar{Q}}$~\cite{bonciani} 
are the following perturbative functions
\ba
A_q(\as) &=& C_F \, \frac{\as}{\pi} \, \left[ 1+ \frac{K}{2} \frac{\as}{\pi}
\right] + {\cal O}( \as^3) \, , \label{afq} \\
A_g(\as) &=& C_A \, \frac{\as}{\pi} \, \left[ 1+ \frac{K}{2} \frac{\as}{\pi}
\right] + {\cal O}( \as^3) \, , \label{afg} \\
D_{Q {\bar Q}}(\as) &=& - C_A \, \frac{\as}{\pi} + {\cal O}( \as^2)\, . 
\label{dfq}
\ea
and the coefficient $K$ is given by:
\be
K= \left(\frac{67}{18} - \frac{\pi^2}{6} \right) C_A - \frac{10}{9} T_R N_f \;.
\ee 
Some comments are in order.

The LL terms $\as^n \ln^{n+1} N$ in eqs.~(\ref{dqq1}-\ref{dgg8}) are obtained
by neglecting the contribution of the function $D_{Q \bar{Q}}$ and by
truncating $A_q(\as)$ and $A_g(\as)$ to their first order. One thus recovers the
LL resummed results first derived in ref.~\cite{Laenen}.                    
 
The contributions of the functions $A_q, A_g, D_{Q \bar{Q}}$ have a direct
physical interpretation. The function $A_q \,(A_g)$ measures soft {\em and}
collinear radiation from the $q{\bar q}$~$(gg)$ partonic channel in the
initial state. This is the only contribution that appears in hadroproduction
processes of colourless heavy-mass systems, such as the DY 
process~\cite{Sterman,CT,CMW} or Higgs boson production via gluon
fusion~\cite{CdET,Spira}.
                         
The function $D_{Q \bar{Q}}$ is due to soft emission from the final-state
heavy-quark pair. Since the final-state quarks are massive, they do not
lead to collinear logarithms and the function $D_{Q \bar{Q}}$ only enters 
at NLL accuracy. At threshold, the heavy-quark pair acts as a single 
final-state
system with invariant-mass squared $Q^2=4m^2$ and with colour
charge given by the total $Q{\bar Q}$ charge. Thus, its contribution
to the radiative factors $\Delta_{ij,{\bf I}, \, N}$ vanishes in the 
colour-singlet channels (see eqs.~(\ref{dqq1},\ref{dgg1}))
and is proportional to $C_A$ and independent of the initial state
in the colour-octet channels (see eqs.~(\ref{dqq8},\ref{dgg8})).

Since we know the resummed radiative factors $\Delta_{ij,{\bf I}, \, N}$
only to NLL accuracy, 
we use  the expressions (\ref{dqq1}-\ref{dgg8})
by explicitly carrying out the $z$ and $q_{\perp}^{2}$ integrals 
and neglecting terms beyond NLL. Thus, we write:
\ba                              
\Delta_{ij,{\bf I}, \,N} \left( 
\as(\mu^2), \mu^2/m^2 \right) &=& \exp \left\{ \ln N
\; g^{(1)}_{ij} ( b_0 \, \as(\mu^2) \ln N) + 
g^{(2)}_{ij,{\bf I}} (b_0 \, \as(\mu^2) \ln N, \mu^2/m^2) 
\right. \nn \\              
&+& \left. {\cal O}(\as^2(\as \ln N)^k) \frac{}{} \right\} \;\;.
\label{dnexp} 
\ea
The functions $g^{(1)}$ and $g^{(2)}$ resum the LL and NLL terms, respectively.
They can be explicitly computed as in refs.~\cite{CT,CMNT2}. We find:
\beq
g^{(1)}_{q{\bar q}}(\lambda) = \frac{C_F}{\pi b_0 \lambda}
\left[ 2\lambda + (1-2\lambda) \ln (1-2\lambda) \right] \;, \quad 
g^{(1)}_{gg}(\lambda) = \frac{C_A}{C_F} \, g^{(1)}_{q{\bar q}}(\lambda)
\;, \label{g1fun}                         
\eeq
and
\ba
g^{(2)}_{q{\bar q},{\bf 1}}(\lambda,\mu^2/m^2) &=&
- \gamma_E \frac{2C_F}{\pi b_0} \ln (1-2\lambda)
+\frac{C_F  b_1}{\pi b_0^3}
\left[ 2\lambda + \ln (1-2\lambda) + \frac{1}{2} \ln^2 (1-2\lambda) \right]
\nn \\
&-& \frac{C_F K}{2\pi^2 b_0^2} \left[2\lambda + \ln (1-2\lambda) \right] 
- \frac{C_F}{\pi b_0} \ln (1-2\lambda) \ln \frac{\mu^2}{4m^2} \;, \nn \\
g^{(2)}_{gg,{\bf 1}}(\lambda,\mu^2/m^2) &=& \frac{C_A}{C_F} \,
g^{(2)}_{q{\bar q},{\bf 1}}(\lambda,\mu^2/m^2) \;, \label{g2fun} \\
g^{(2)}_{ij,{\bf 8}}(\lambda,\mu^2/m^2) &=&
g^{(2)}_{ij,{\bf 1}}(\lambda,\mu^2/m^2) 
- \frac{C_A}{2\pi b_0} \ln (1-2\lambda) \;. \nn
\ea
where $b_0, \,b_1$ are the first two coefficients of the QCD $\beta$-function
\beq
\label{betacoef}
b_0 = \frac{11C_A - 4T_R N_f}{12 \pi} \;, \quad 
b_1 = \frac{17C_A^2 - 10 C_A T_R N_f - 6 C_F T_R N_f}{24 \pi^2} \;\;.
\eeq
Note that the LL function $g^{(1)}_{ij, \,N}$ depends 
on the initial-state partonic channel, but it is independent of the colour
state.

It is common practice to study the scale dependence of the cross-section
in order to assess the impact of neglected higher-order effects. 
Normally, in the fixed-order calculations,        
one considers the range $m/2<\mu<2m$. When the threshold region is
approached, however, a new scale comes into play, namely the
typical distance from threshold of the hard production process.
Thus, the process is really a two scale process, and
one is forced to use a wider scale range in order to estimate the error.
When the NLO calculation is improved by LL resummation, this is no longer
needed, since the large threshold logarithms have been properly resummed.
Thus, one can still safely consider the range $m/2<\mu<2m$.
Observe, however, that the inclusion of LL resummation does not
bring a reduction of the scale dependence in the range $m/2<\mu<2m$.
The improvement comes simply from the fact that in this case
we are allowed to use scales
of order $m$. On the other hand, when NLL resummation
is used, one does expect a reduction of the scale dependence in the
range $m/2<\mu<2m$.                                       
This can be seen explicitly from eqs.~(\ref{dnexp}-\ref{g2fun}).
A scale variation from $\mu$ to $\mu^\prime$ is perturbatively
equivalent to the replacement $\as(\mu^2) \to \as(\mu^{\prime 2}) =
\as(\mu^2)[1 - b_0 \as(\mu) \ln \mu^{\prime 2}/\mu^2 + \dots]$. This
replacement in the LL functions $g^{(1)}_{ij}(b_0 \as \ln N)$ leads
to additional NLL terms in the exponent on the right-hand side of 
eq.~(\ref{dnexp}). These additional terms are partly cancelled by those
that originate from the explicit $\mu$-dependence of the functions
$g^{(2)}_{ij,{\bf I}}(b_0 \as \ln N,\mu^2/m^2)$. The left-over
NLL contributions exactly match the scale dependence of the parton densities
at large values of $N$. For instance, in the $q{\bar q}$ channel we have
\beq
\Delta_{q{\bar q},{\bf I}, \,N} 
\left( \as(\mu^{\prime 2}), \frac{\mu^{\prime 2}}{m^2} \right) 
= \Delta_{q{\bar q},{\bf I}, \,N}
\left( \as(\mu^2), \frac{\mu^2}{m^2} \right)
\exp \left\{ 2 \frac{C_F}{\pi} \as(\mu^2) \ln N \ln \frac{\mu^{\prime 2}}{\mu^2}
+ {\cal O}(\as(\as \ln N)^k) \frac{}{} \right\} \;,
\eeq
and, in the calculation of the full hadronic cross-section,
the exponential on the right-hand side is compensated by the scale dependence
of the quark densities in the $\MSB$ factorization scheme:
\beq
F_{q,N}(\mu^{\prime 2}) F_{{\bar q},N}(\mu^{\prime 2})
= F_{q,N}(\mu^2) F_{{\bar q},N}(\mu^2)
\exp \left\{ - 2 \frac{C_F}{\pi} \as(\mu^2) \ln N 
\ln \frac{\mu^{\prime 2}}{\mu^2}
+ {\cal O}(\as(\as \ln N)^k) \frac{}{} \right\} \;.
\eeq          

\subsection{NLL resummed cross-section}
Using the results of Sect.~\ref{radfact}, we can introduce the NLL resummed
partonic cross-sections as follows                             
\beq                    
\label{fijress} 
{\hat \sigma}_{ij, \, N}^{(res)}(m^2,\as(\mu^2),\mu^2) = 
\frac{\as^2(\mu^2)}{m^2} \sum_{{\bf I = 1,8}}
f_{ij,{\bf I}, \, N}^{(res)}(\as(\mu^2),\mu^2/m^2) \;\;, 
\eeq
\beq
\label{fijres}
f_{ij,{\bf I}, \, N}^{(res)}(\as(\mu^2),\mu^2/m^2) =
f_{ij,{\bf I}, \, N}^{(corr)}(\as(\mu^2),\mu^2/m^2) \;
\Delta_{ij,{\bf I}, \,N+1}\!\left( \as(\mu^2), \frac{\mu^2}{m^2} \right)
\,.
\eeq
Note the mismatch between the moment indices in the right-hand side 
of eq.~(\ref{fijres}).                          
The radiative factors $\Delta_{ij,{\bf I}, \,N+1}$ depend on the moment 
index $N+1$, like the parton densities in the factorization formula
(\ref{HVQfN}).

The terms $f_{ij,{\bf I}, \, N}^{(corr)}$ in eq.~(\ref{fijres}) are
given by the the LO functions $f_{ij,{\bf I}}^{(0)}$ in 
eqs.~(\ref{fij018}-\ref{fgg01}) after correction by the Coulomb contributions 
in eqs.~(\ref{f1qqC},\ref{f1ggOC},\ref{f1ggSC}) and by the $N$-independent
coefficients $C_{ij}$ in eqs.~(\ref{coefqq},\ref{coefgg})
\be
\label{fcorr}
f^{(corr)}_{ij,{\bf I}, \,N} = 
\left( f^{(0)}_{ij,{\bf I}, \,N} + 
4 \pi \as(\mu^2) f_{ij,{\bf I}, \,N}^{(1),{\rm Coul}} \right) \left[1 +
\frac{\as(\mu^2)}{\pi} C_{ij}(\mu^2/m^2) \right] \;\;.
\ee

Using the definition in eq.~(\ref{fcorr}) and
the perturbative expansions of the radiative factors 
$\Delta_{ij,{\bf I}, \,N}$,
\beqn
\ln \Delta_{q{\bar q},{\bf 1}, \,N} &=& \frac{C_F}{C_A} 
\ln \Delta_{gg,{\bf 1}, \,N} = \frac{\as(\mu^2)}{\pi} 
\left[ 2 C_F \ln^2 N + (4C_F \gamma_E + 2 C_F \ln \mu^2/m^2) \ln N \right]
+ {\cal O}(\as^2) \;, \nn \\
\ln \Delta_{ij,{\bf 8}, \,N} &=& \ln \Delta_{ij,{\bf 1}, \,N} +
\frac{\as(\mu^2)}{\pi} C_A \ln N + {\cal O}(\as^2) \;,
\eeqn
it is straightforward to check that eqs.~(\ref{fijress},\ref{fijres})
correctly reproduce the NLO
threshold behaviour in eqs.~(\ref{f1qqN},\ref{f1ggN}). In particular, since
both the singlet and octet states are produced at LO via $gg$ fusion, the
colour factor $(N_c^2-4)/(N_c^2-2)$ that appears in eq.~(\ref{f1ggN}) is due
to the fact that final-state soft-gluon radiation is weighted by
the ratio
\beq
\frac{f_{gg,{\bf 8}, \,N}^{(0)}}{f_{gg,N}^{(0)}} = \frac{N_c^2-4}{N_c^2-2} 
\;\left( 1 + {\cal O}(1/N) \right) \;,
\eeq
which differs from unity near threshold.

The all-order factorization of $f^{(corr)}_{ij,{\bf I}, \,N}$ with respect to 
the radiative factors $\Delta_{ij,{\bf I}, \,N+1}$ in the resummed
partonic cross-sections (\ref{fijres}) is justified by the fact that 
the ${\cal O}(\as)$-terms in eq.~(\ref{fcorr}) are produced by non-soft {\em
virtual} corrections to the LO subprocesses. The resummation of Coulomb 
corrections was considered in ref.~\cite{cmnt21} and it turned out that,
in the experimental configurations of practical interest,
its quantitative effect is much smaller than that due to soft-gluon radiation.
Because of this reason, in our resummed formulae we do not include pure Coulomb 
effects beyond NLO, but simply the soft-gluon corrections to the $\oacube$
Coulomb contribution.
                     
Note, however, that we include the constant terms $C_{ij}(\mu^2/m^2)$ as in 
eqs.~(\ref{fijres},\ref{fcorr}). This procedure is analogous to that used
in NLL resummed predictions for other observables, such as $e^+e^-$ event
shapes~\cite{CTTW}. As shown in Sect.~\ref{plres},
these constant terms are important to accurately match the exact near-threshold
behavior at NLO. Moreover, when combined with the NLL soft-gluon factors
$\Delta_{ij,{\bf I}, \,N}$, they lead to the following expansion
\beqn
\left[1 +
\frac{\as(\mu^2)}{\pi} C_{ij}(\mu^2/m^2) \right] \; 
\Delta_{ij,{\bf I}, \,N}(\as(\mu^2), \mu^2/m^2)
\!\!\!&=&\!\!\! 1 + \sum_{n=1}^{+\infty} \as^n(\mu^2) 
\left[ c_{n,2n} \ln^{2n} N 
+ c_{n,2n-1} \ln^{2n-1} N \right. \nn \\
\!\!\!&+&\!\!\! \left. c_{n,2n-2}(\mu^2/m^2) \ln^{2n-2} N
+ {\cal O}(\ln^{2n-3} N) \right] \;,
\label{eq:nnll}
\eeqn 
that correctly predicts the value of the coefficients $c_{n,2n-2}(\mu^2/m^2)$, 
as well as of $c_{n,2n}$ and $c_{n,2n-1}$ that are independent of $C_{ij}$. Note
also that, while $c_{n,2n}$ and $c_{n,2n-1}$ are $\mu$-independent, the
coefficients $c_{n,2n-2}(\mu^2/m^2)$ are not. Their dependence on $\mu$ is
obtained by combining that of $C_{ij}(\mu^2/m^2)$ with the explicit scale
dependence of $\Delta_{ij,{\bf I}, \,N}(\as(\mu^2), \mu^2/m^2)$ at NLL order.
                                                                             
Including the $C_{ij}$ constant term can also be viewed as using the known
NLO cross-section to improve the resummation formula beyond
the NLL approximation. This can be seen schematically in the following
way. We write the left-hand side of eq.~(\ref{eq:nnll}) as:
\beq                                                 
\ln\left[(1+\as/\pi C) \Delta \right]=\ln\Delta + \as/\pi C\,,
\eeq
and since our NLL expression for $\ln\Delta$ has the form
\beq                             
\ln\Delta=\ln N g^{(1)}(\as\ln N) + g^{(2)}(\as\ln N)\,,
\eeq
according to eq.~(\ref{deltadyex}), the term $\as/\pi\, C$ corresponds
precisely to the $\oas$ term in the expansion of the NNLL contribution
$\as g^{(3)}(\as\ln N)$.

We use soft-gluon resummation to NLL accuracy 
to introduce an improved hadronic cross-section $\sigma_N^{(res)}$ as follows
\beqn 
\sigma_N^{(res)}(m^2) &=& \sum_{ij=q{\bar q},gg}  
\;F_{i,N+1}(\mu^2) \;F_{j,N+1}(\mu^2)
\left[ {\hat \sigma}_{ij,N}^{(res)}(m^2, \as(\mu^2),\mu^2)
- \left( {\hat \sigma}_{ij,N}^{(res)}(m^2, \as(\mu^2),\mu^2)
\right)_{\as^3} \right] \nn \\
\label{HQCSNres}
&+& \sigma_N^{(NLO)}(m^2) \;\;,
\eeqn
where $\sigma_N^{(NLO)}$ is the hadronic cross-section at NLO,
${\hat \sigma}_{ij,N}^{(res)}$ is given in eq.~(\ref{fijress}) and
$\left( {\hat \sigma}_{ij,N}^{(res)} \right)_{\as^3}$ represents its
perturbative truncation at order $\as^3$. Thus, because of the subtraction
in the square bracket on the right-hand side, eq.~(\ref{HQCSNres}) exactly
reproduces the NLO results and resums soft-gluon effects beyond
${\cal O}(\as^4)$ to NLL accuracy. This defines our NLO+NLL predictions.        
                                 
The resummed formulae presented so far are given in $N$-moment space.
To obtain cross-sections in the physical $x$-space, one has to perform
the inverse Mellin transformation:
\beq \label{MPHQ}
\sigma^{(\rm res)}(\rho,m^2) =\frac{1}{2\pi i}
\int_{C_{\rm MP}-i\infty}^{C_{\rm MP}+i\infty}\; dN \;\rho^{-N}
\sigma_N^{(\rm res)}(m^2) \;.        
\eeq

When the $N$-moments $\sigma_N$ are evaluated at a fixed perturbative order
in $\as$, they are analytic functions in a
right half-plane of the complex variable $N$. In this case, the constant $C_{\rm
MP}$
that defines the integration contour in eq.~(\ref{MPHQ}) has to be chosen in
this half-plane, i.e. on the right of all the possible singularities of the 
$N$-moments.

An additional complication occurs when the $N$-moments are computed in resummed
perturbation theory. In this case, since the resummed functions
$g_{ij,{\bf I}}^{(k)}(\lambda)$ in eqs.~(\ref{g1fun},\ref{g2fun})
are singular at $\lambda=1/2$, the soft-gluon factors  
$\Delta_{ij,{\bf I}, \,N}(\as(\mu^2),\mu^2/m^2)$ in eq.~(\ref{dnexp})
have cut singularities that start at the branch-point $N=N_L=\exp(1/2b_0\as)$. 
These singularities, which are produced in eqs.~(\ref{dqq1}-\ref{dgg8}) by
the $q_{\perp}$-integration down to the
Landau pole of the running coupling $\as(q_{\perp}^2)$, signal the onset of
non-perturbative phenomena at very large values of $N$ or, equivalently,
in the region very close to threshold.  
 
The issue of how to deal with the Landau singularity in soft-gluon
resummation formulae for hadronic collisions was discussed in detail in 
ref.~\cite{CMNT2}. In the evaluation of the inverse Mellin transformation
(\ref{MPHQ}) we thus use the {\em Minimal Prescription} introduced in 
ref.~ \cite{CMNT2}. The constant $C_{\rm MP}$ is chosen in such a way that all
singularities in the integrand are to the left of the integration contour,
except for the Landau singularity at $N=N_L$, which should lie to the far right.
This prescription is consistent~\cite{CMNT2} with the perturbative content of 
the soft-gluon resummation formulae because it converges asymptotically to the
perturbative series and it does not introduce (unjustified) power corrections
of non-perturbative origin. These corrections are certainly present in physical
cross-sections, but their effect is not expected to be sizeable
as long as $m$ is sufficiently perturbative and $\rho$ is
sufficiently far from the hadronic threshold. Obviously, approaching the
essentially non-perturbative regime $m \to 1 \,{\rm GeV}, \rho \to 1$,
a physically motivated treatment of non-perturbative effects has to be
introduced.
In the following Section, we limit ourselves to present
numerical and phenomenological results for kinematic configurations in which
non-perturbative corrections
should be smaller than the estimated uncertainty of the perturbative
predictions.

In all cases considered, we have verified
that the minimal prescription is in practice equivalent to evaluate eq.~(\ref{MPHQ})
order by order in perturbation theory, until the corrections become
numerically insignificant. An example of this procedure will also be shown in
Section~\ref{HLResults}.
                       
\section{Results}
\label{Results}
We present in this Section some numerical results, to provide an illustration
of the size of the effects considered and a new estimate of heavy-quark 
production cross-sections of relevance for present and future hadronic
colliders.
\subsection{Parton-level results}
\label{plres}
\begin{figure}
\begin{center}
\centerline{
\epsfig{file=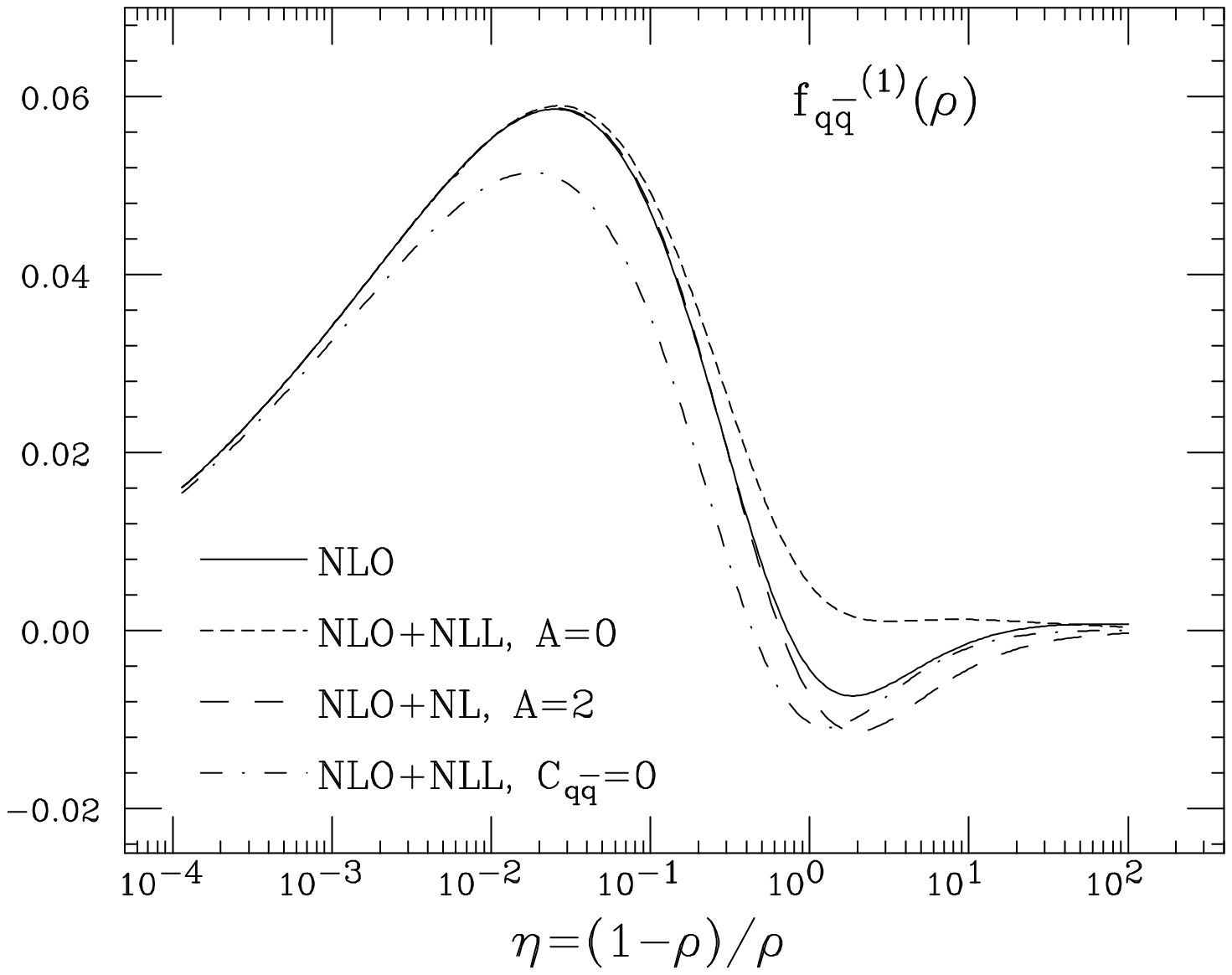,width=0.50\textwidth,clip=}\hfil
\epsfig{file=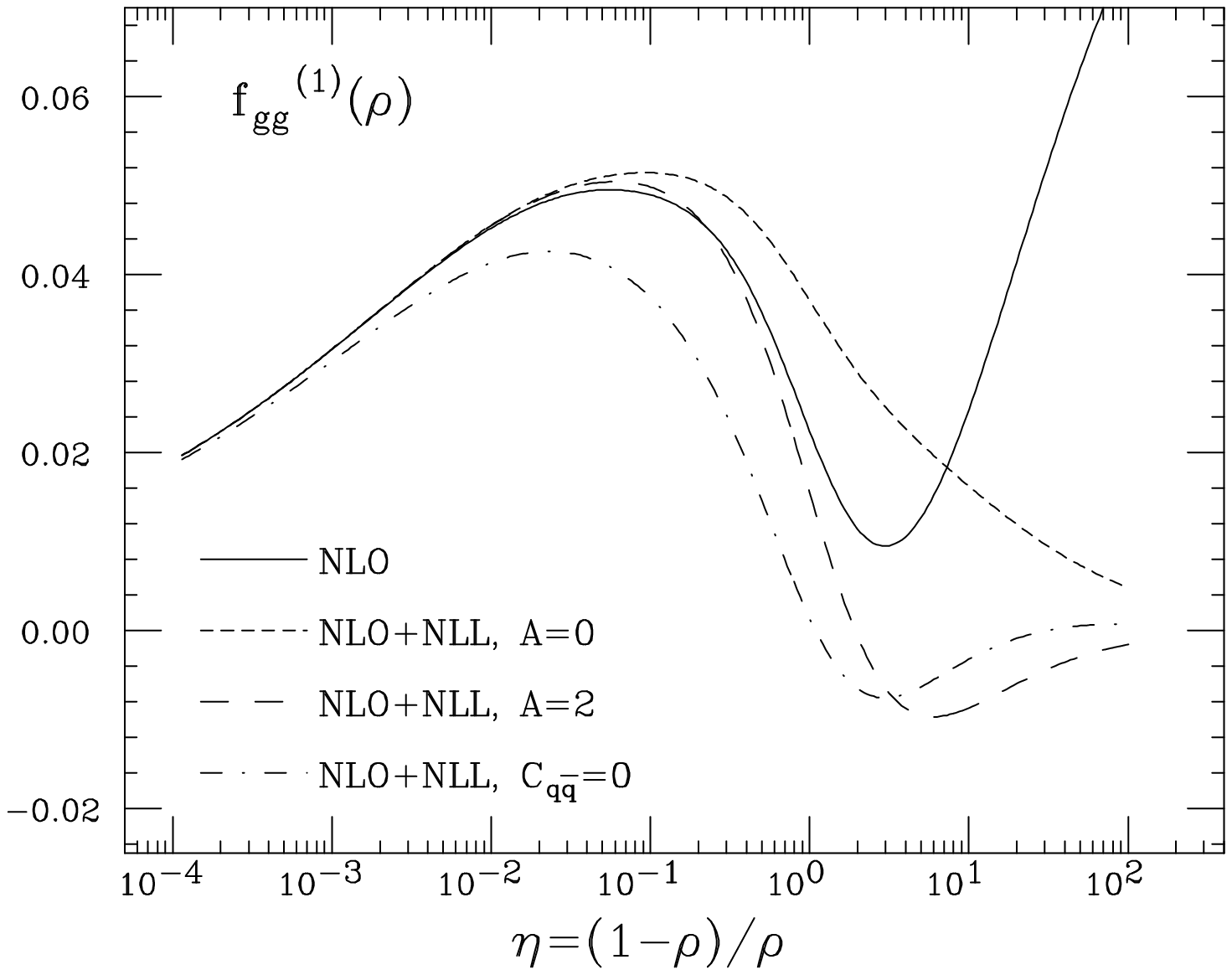,width=0.50\textwidth,clip=}                
}                                
\ccaption{}{\label{fig:qqnlo} 
Left (Right):
the function $f_{q\bar q}^{(1)}(\rho)$ ($f_{gg}^{(1)}(\rho)$), 
plotted as a function of                      
$\eta=(1-\rho)/\rho$. The solid line represents the exact NLO
result~\cite{NDE,VN}; the short-dashed line corresponds to the $\oacube$
truncation of the resummed result defined by 
eqs.~(\ref{fijress}--\ref{fcorr}); the dot-dashed line is obtained 
from this last
result by setting the constant $C_{q\bar q}$ ($C_{gg}$) to 0; 
the dashed line is obtained instead by the                 
replacement in eq.~(\ref{eq:Cshift}), with $A=2$.}
\end{center}
\end{figure}
     
 
We start by discussing the resummation effects at the level of partonic cross
sections. The resummed partonic cross-section for the production of a heavy
quark pair can be obtained from eqs.~(\ref{HQCSNres})
and (\ref{MPHQ}) by assuming structure    
functions of the form $F(x)=\delta(1-x)$, and therefore $F_N=1$ for all complex
values of $N$. 

We consider first the $\oacube$ terms in the expansion of the resummed
cross-section, in order to estimate to which accuracy this reproduces the
exact NLO results.                                                  
In fig.~\ref{fig:qqnlo} (left) we plot the function $f^{(1)}_{q\bar q}$, defined
in eq.~(\ref{fpert}),  as a function of $\eta=(1-\rho)/\rho$.
The exact $\oacube$ result~\cite{NDE} is compared with
three possible implementations
of the resummation procedure, all equivalent at NLL. 
In one case (dot-dashed line) we set the constant $C_{q\bar q}$ introduced in
eqs.(\ref{coefqq}) and (\ref{fcorr}) equal to 0.    
In the second case (short-dashed line) we include the contribution of 
$C_{q\bar q}$. In a
third case we correct the contribution of the constant $C_{q\bar q}$ 
by a term which is
suppressed by a factor of $1/N$, which does not introduce poles on the real $N$
axis,  and which gives vanishing first moment to $f^{(1)}_{q\bar q,N}$:
\be \label{eq:Cshift}                            
      C_{ij} \to C_{ij}\; (1-\frac{A}{N+A-1}) \; , \qquad ij=q\bar q, gg \; .
\ee                                                
In our applications we shall consider the two cases with $A=0$ (namely no
correction to the $C_{ij}$ term) and $A=2$ as a way to establish the size of
subleading corrections beyond the NLL order.           
In all cases we include the effect of the leading-order Coulomb effects, as
described in eq.~(\ref{fcorr}).
                               
\begin{figure}
\begin{center}
\centerline{
\epsfig{file=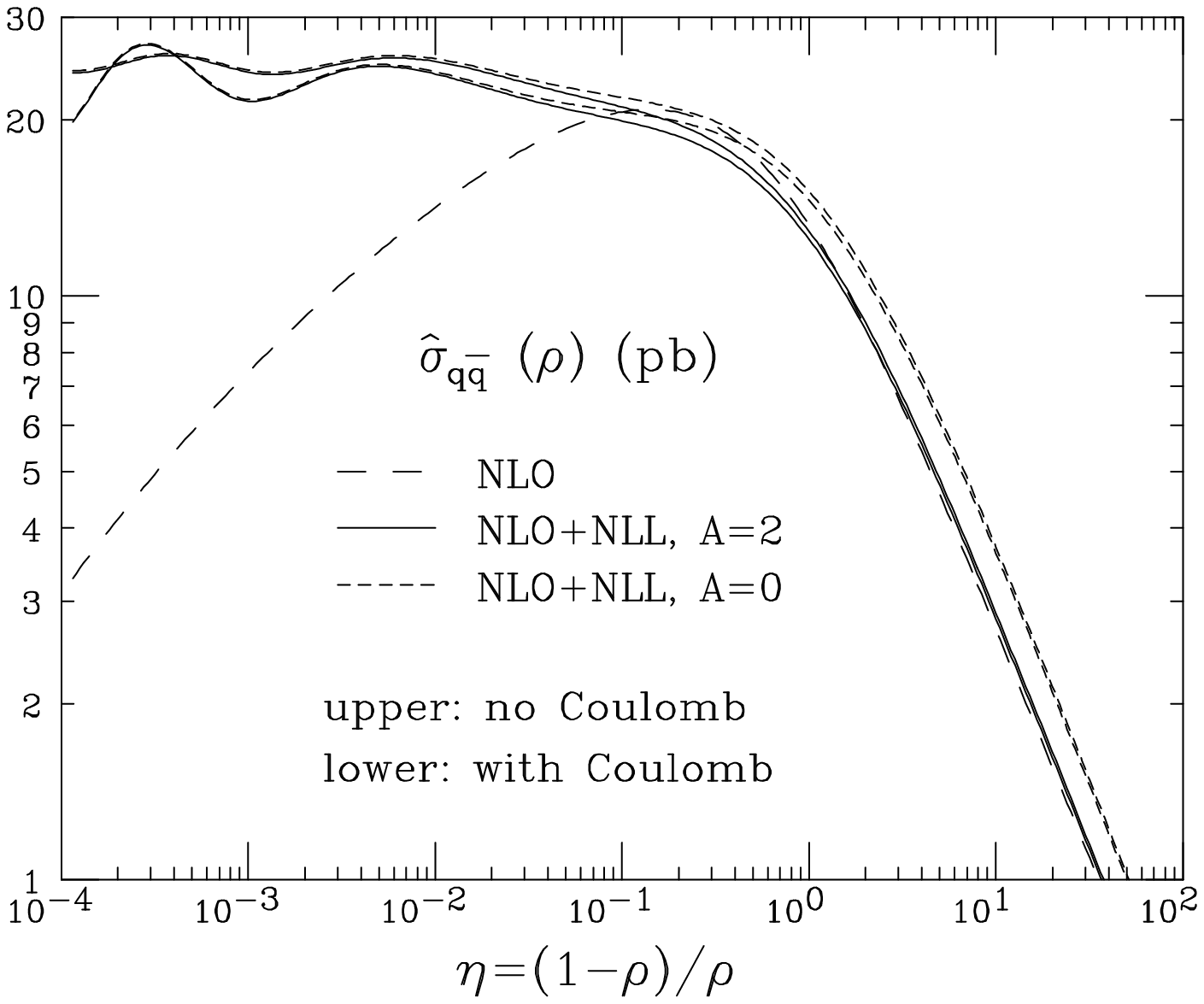,width=0.50\textwidth,clip=}\hfil
\epsfig{file=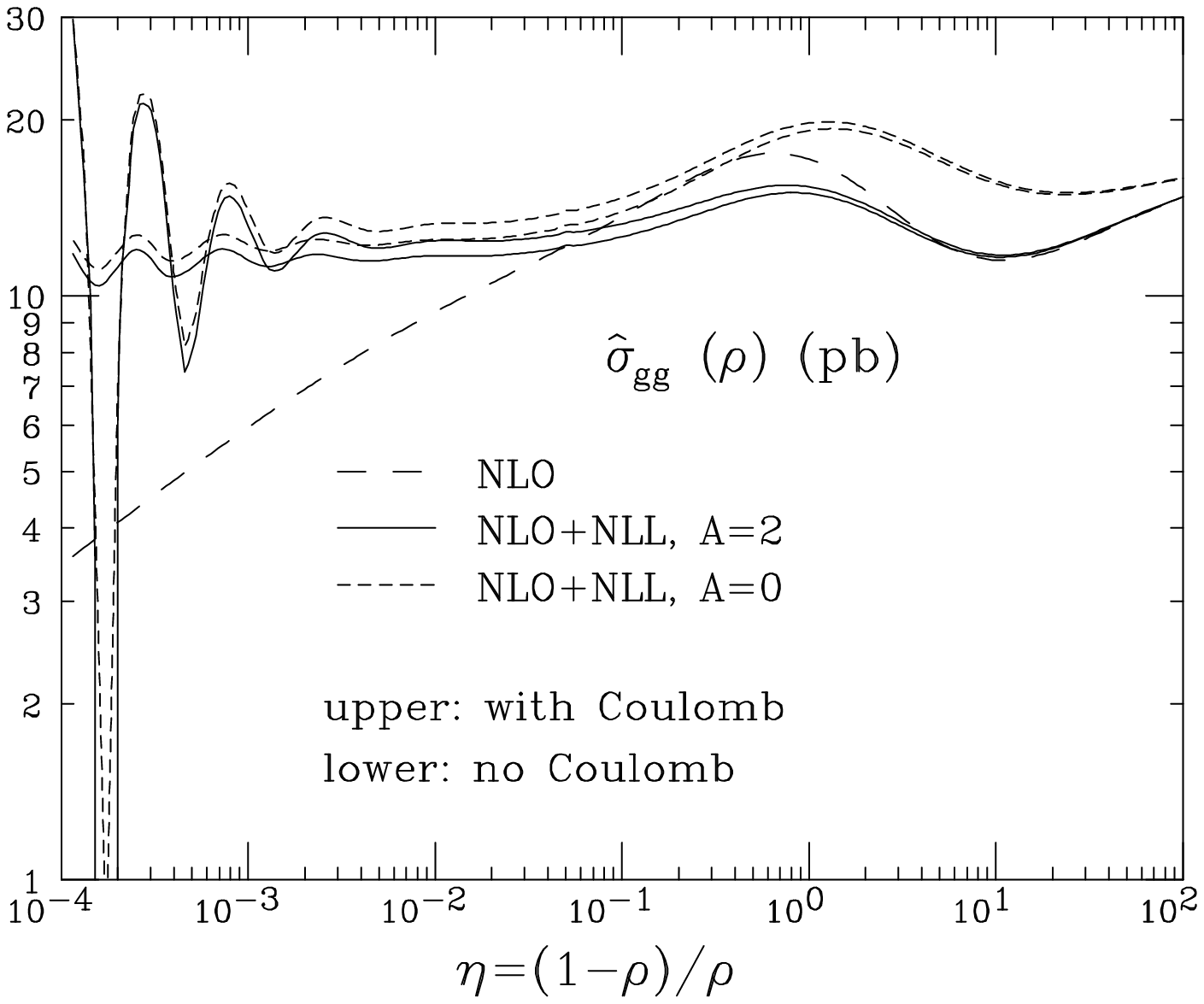,width=0.50\textwidth,clip=}
}                                
\ccaption{}{\label{fig:qqres} 
Partonic cross-section for the processes $q\bar q \to Q \overline{Q}$  (left)
and $gg \to Q \overline{Q}$  
(right) (in pb,                                                    
and for $m_{Q}=175$~GeV). The dashed line is the exact NLO
result~\cite{NDE,VN};  the short-dashed (solid) lines correspond to the NLO+NLL
result, with the coefficient $A$ defined in eq.~(\ref{eq:Cshift}) equal to 0
(2). The lower and upper curves correspond to  inclusion or neglect of the
Coulomb contribution.
}
\end{center}
\end{figure}

As one can see from fig.~\ref{fig:qqnlo}, the inclusion of the finite term
$C_{q\bar{q}}$                                                       
is essential to accurately reproduce shape and normalization of the function 
$f^{(1)}_{q\bar q}$ near threshold. The agreement deteriorates unavoidably for
$\rho \gg 1$, as here terms subleading in $1/N$ become important.
Note that the
two choices $A=0$ and $A=2$ braket the exact result, and thus provide a good
estimator of the subleading terms' systematics.
The choice $A=2$, furthermore, provides a very accurate description up to
values of
$\eta$ of order 1. This is the region which dominates the production 
cross-section in the cases of interest.
      
Analogous results for the $gg$ channel are given 
in the right panel of fig.~\ref{fig:qqnlo}.
The agreement is again very good near threshold. Far above threshold the exact
NLO result is dominated by the $t$-channel gluon exchange diagrams~\cite{NDE,
smallx},
which give rise to a $1/N$ pole not controlled by the soft-gluon
resummation.

The fully resummed parton-level cross-sections are shown in
fig.~\ref{fig:qqres} for the $q\bar q$ and $gg$ channels (left and right
panel, respectively). Here and in the following we shall                        
define the resummed cross-sections as in eq.~(\ref{HQCSNres}), that is, we
substitute their $\oacube$ terms with                                 
the exact NLO result,
using the same choice of renormalization scale $\mu$.
In this way our results are exact up to (and including)
$\oacube$, and include the NLL resummation of terms of $\oafour$ and higher.
We compare the fixed-order results (dashed lines) with the resummed results.
For these we provide both the $A=0$ and $A=2$ prescriptions, as well as the
inclusion and neglect of the Coulomb terms in eq.~(\ref{fcorr}).
Note that even at the level of resummed cross-sections the      
prescriptions $A=0$ and $A=2$ braket the fixed order result. Note furthermore
that, for the curves without the Coulomb contribution,
the oscillations of the cross-section near threshold are significantly reduced 
relative to the LL resummation which was presented in fig.~3
of ref.~\cite{CMNT2}.
   
\begin{figure}
\begin{center}
\epsfig{file=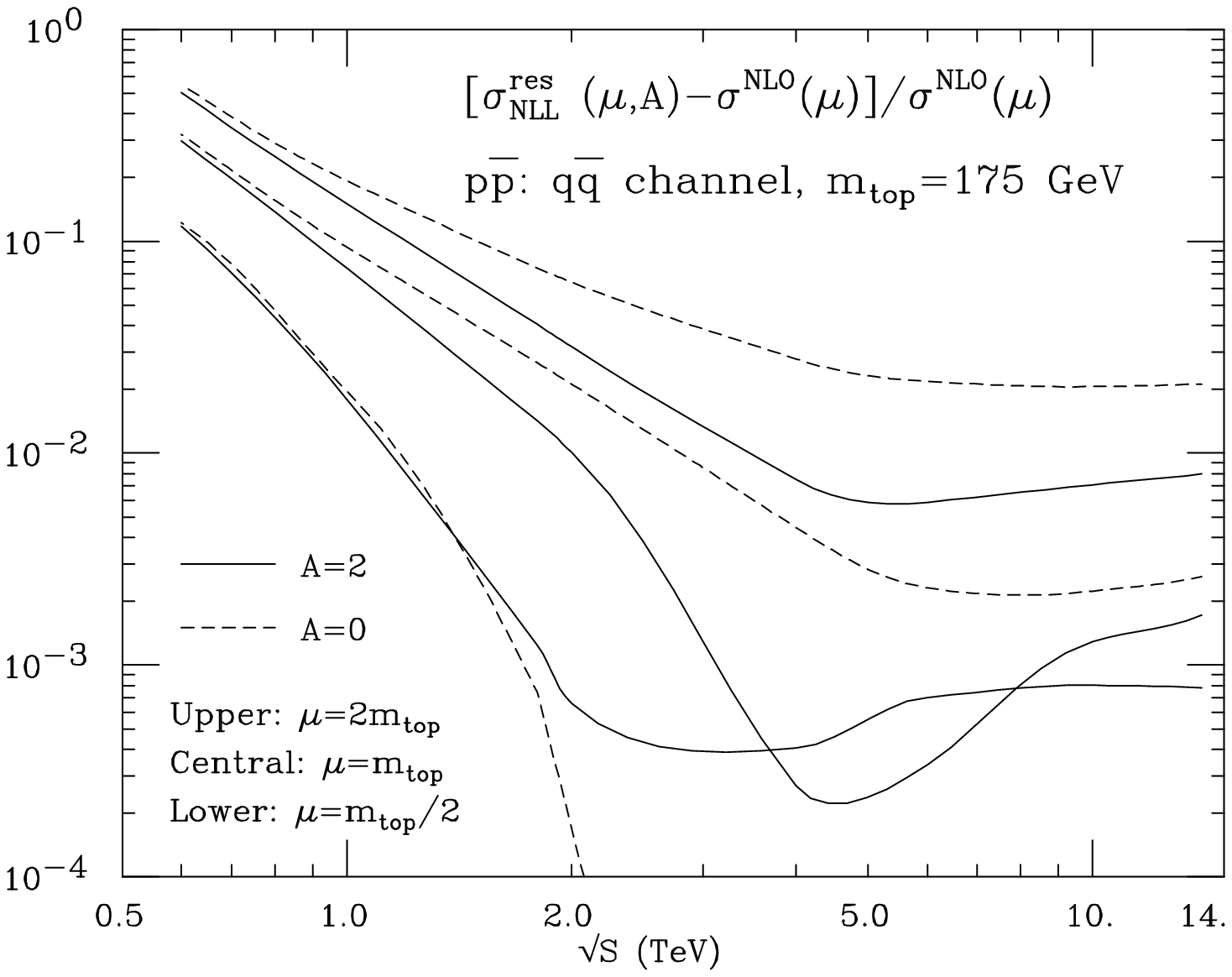,width=0.8\textwidth,clip=}
\vspace*{\mysep}
\ccaption{}{\label{fig:deltaqq} 
Contribution of gluon resummation at order $\oafour$ and higher, relative to
the exact NLO result, for top-pair production via $q\bar q$ annihilation in
$p\bar p$ collisions. The solid (dashed) lines correspond to $A=2$ ($A=0$).
The three sets of curves correspond to the choice of scale $\mu=2m_{t},\;
m_{t}$ and $m_{t}/2$, in {\em descending} order, with
$m_{t}=175$~GeV, and PDF set MRSR2.}
\end{center}                   
\vspace*{\mysep}
\end{figure}
\begin{figure}
\begin{center}
\epsfig{file=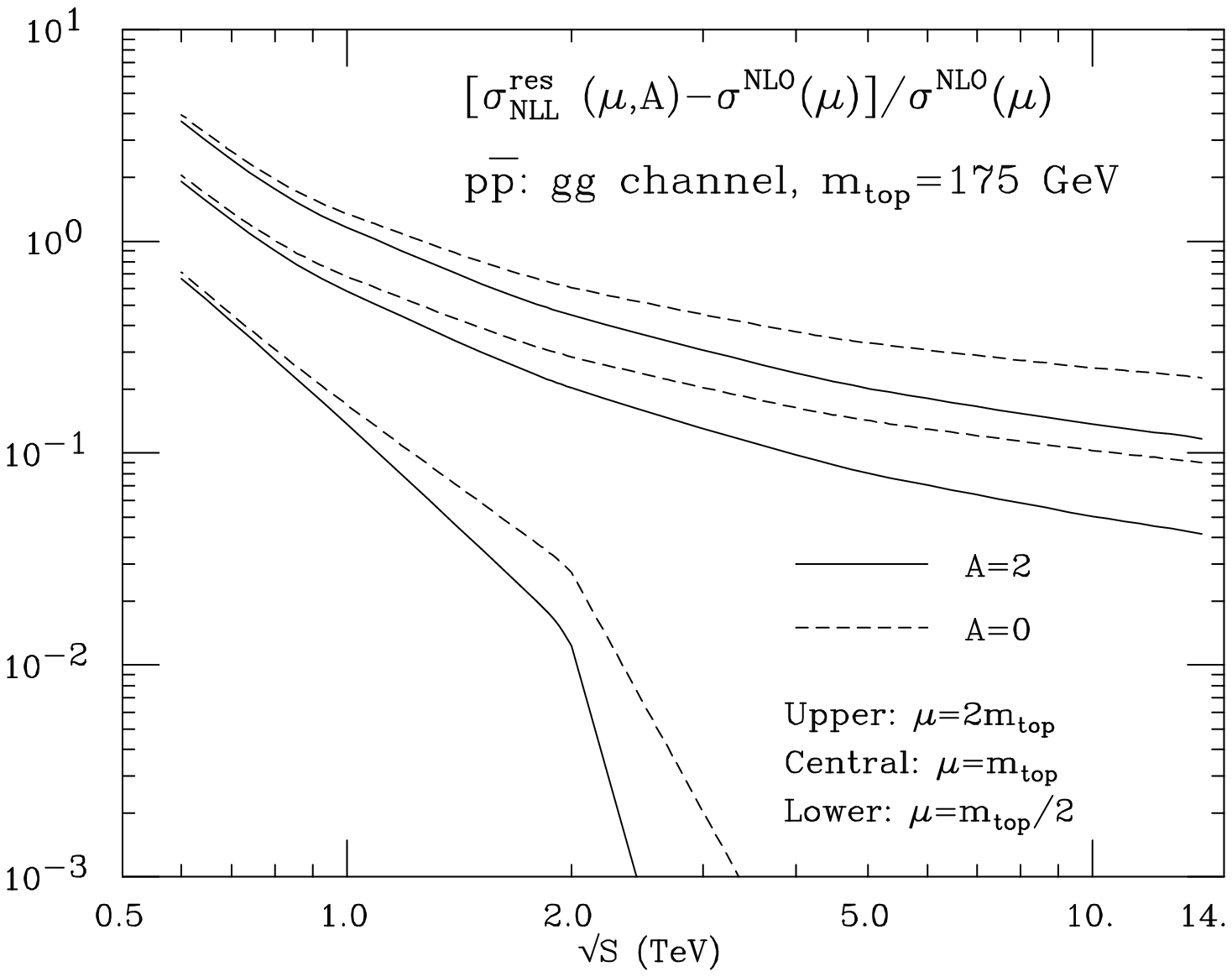,width=0.8\textwidth,clip=}
\vspace*{\mysep}
\ccaption{}{\label{fig:deltagg} 
Same as fig.~\ref{fig:deltaqq}, for 
production via $gg$ annihilation. }
\end{center}
\end{figure}

\begin{figure}
\begin{center}
\epsfig{file=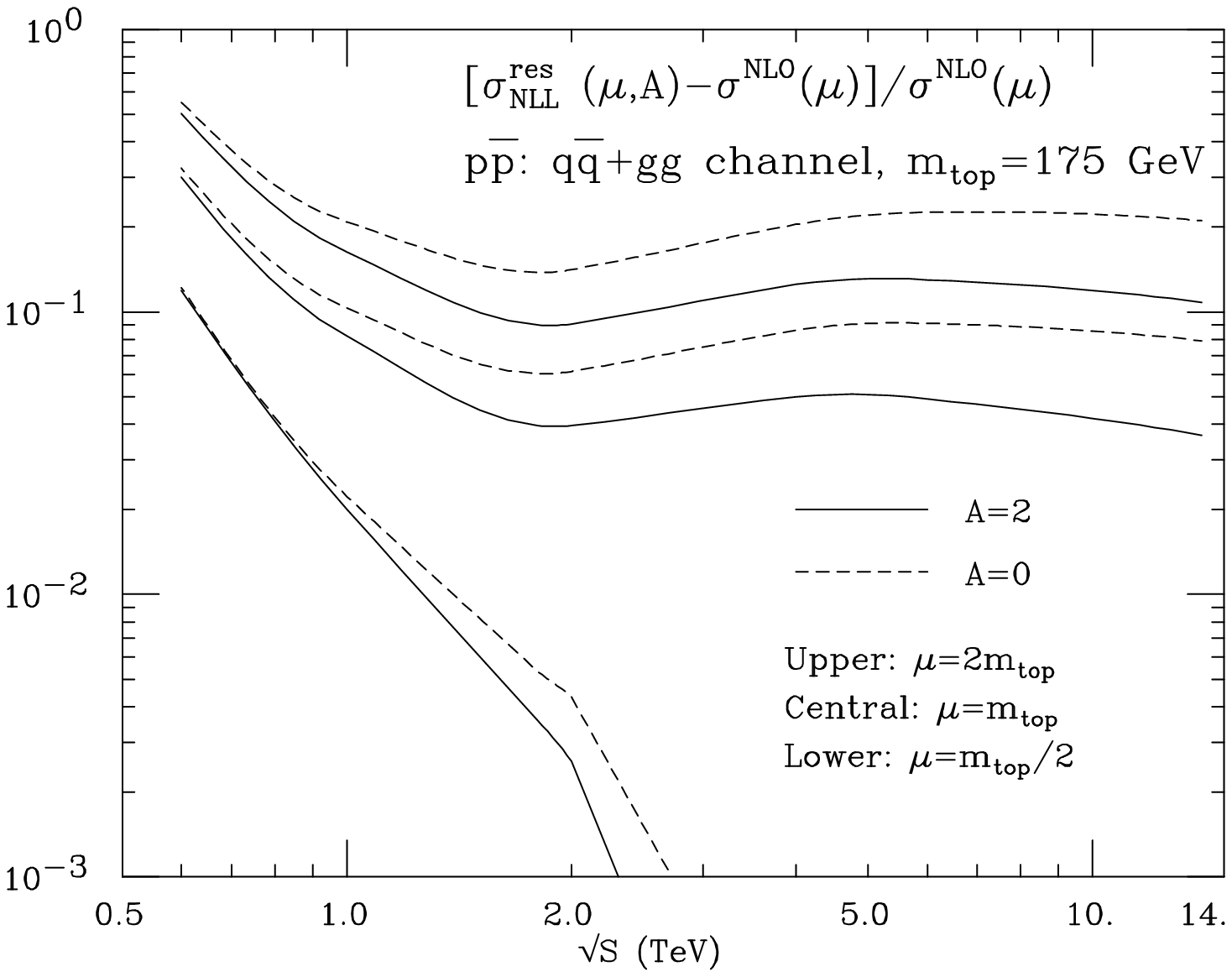,width=0.8\textwidth,clip=}
\vspace*{\hmysep}
\ccaption{}{\label{fig:deltatot} 
Same as fig.~\ref{fig:deltaqq}, for 
the combined production channels $gg+q\bar q$.}
\end{center}
\end{figure}

\subsection{Hadron-level results}
\label{HLResults}
In this Section we present results for the full hadronic cross-sections.
As a default set of parton densities, we shall use the MRSR2 set described
in~\cite{MRSR}. For the top-quark mass we shall use, unless otherwise
indicated, $m_t=175$~GeV.           

Figures \ref{fig:deltaqq}, \ref{fig:deltagg} and \ref{fig:deltatot}
present the ratios:
\beq \label{deltadef} 
    \frac{\sigma_{{\rm NLL,gg}}^{\rm res} - \sigma_{{\rm gg}}^{\rm NLO}}
         {\sigma_{{\rm gg}}^{\rm NLO}}
    \,,\quad 
    \frac{\sigma_{{\rm NLL,q\bar{q}}}^{\rm res} 
         - \sigma_{{\rm q\bar{q}}}^{\rm NLO}}
         {\sigma_{{\rm q\bar{q}}}^{\rm NLO}}
    \,,\quad                   
    \frac{\sigma_{{\rm NLL,(gg+q\bar{q})}}^{\rm res} 
         - \sigma_{{\rm (gg+q\bar{q})}}^{\rm NLO}}
         {\sigma_{{\rm (gg+q\bar{q})}}^{\rm NLO}}
  \, .                              
\eeq
For each channel we present the results using both the $A=0$ (dashed lines) and
$A=2$ (solid lines) prescriptions. We also 
show the dependence on
the choice of renormalization and factorization scales, which we always take
equal, and varying within the set $\mu=(m_t/2,m_t,2m_t)$.
Note that the size of the resummation effects is larger for the larger
scales, contrary to the behaviour of the scale dependence of the NLO cross
section. This suggests that the scale dependence of the resummed cross-section
will be reduced relative to that of the NLO results. 
The size of the resummation corrections is large at small centre-of-mass
energies, as
should be expected for the region where production is dominated by the
threshold region. At Tevatron energies ($\sqrt{S}=1.8$ TeV), the size of the
corrections equals 4\% (6\%) for $A=2$ ($A=0$) and $\mu=m_t$. It becomes equal
to 0.4\% (0.5\%) for $\mu=m_t/2$ and to 9\% (14\%) for $\mu=2m_t$. 

\begin{figure}
\begin{center}
\epsfig{file=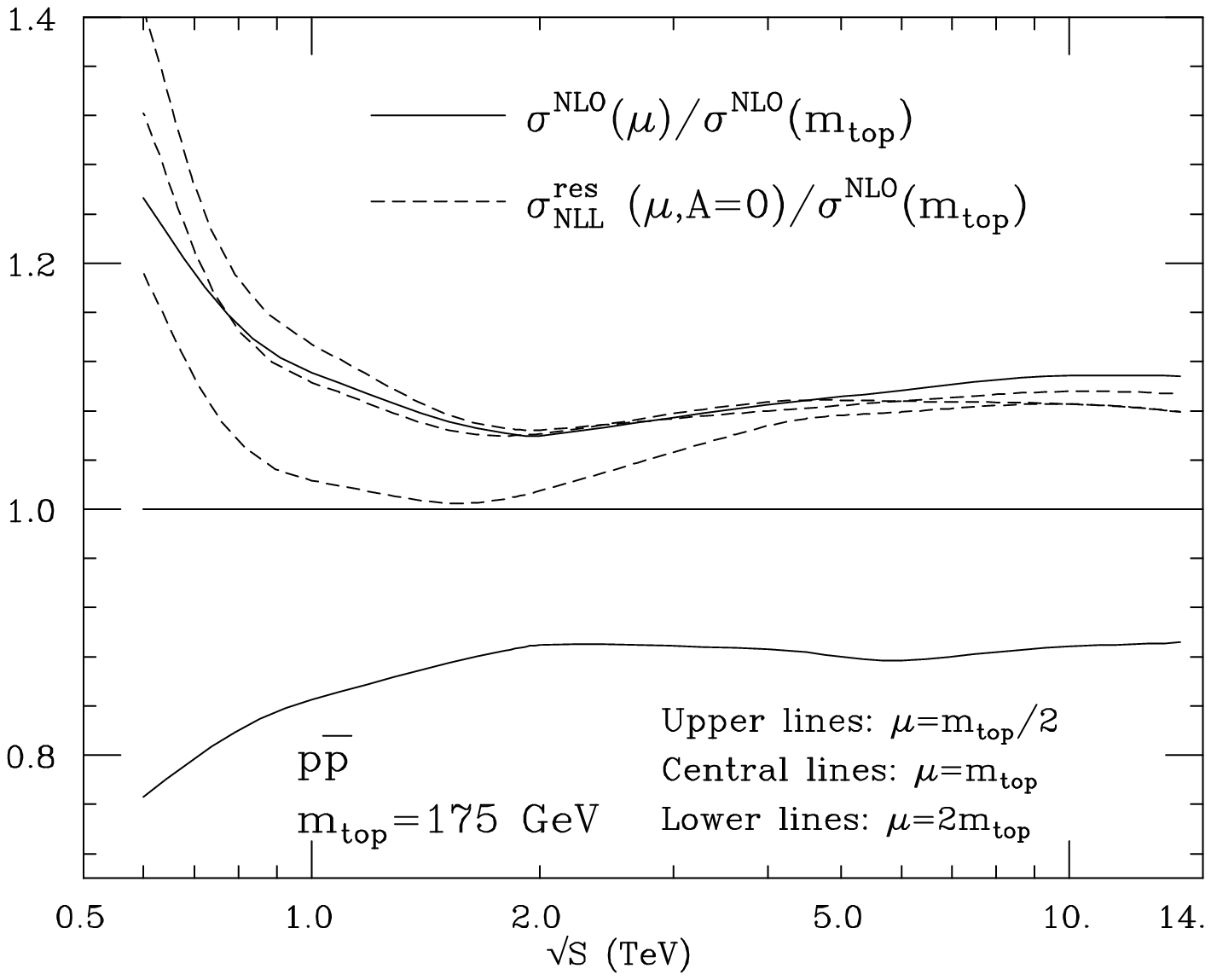,width=0.8\textwidth,clip=}
\vspace*{\mysep}
\ccaption{}{\label{fig:scaleA0} 
Scale-dependence of the $t\bar t$ production cross-section in
$p\bar p$ collisions, as a function of $\sqrt{S}$.
The solid lines represent the exact NLO result for different choices of $\mu$
($\mu=m_{t}/2$ and $2m_{t}$), normalised to the $\mu=m_{t}$ result.
The solid lines represent the NLO+NLL result (with $A=0$) 
for different choices of $\mu$
($\mu=m_{t}/2,\; m_{t}$ and $2m_{t}$), 
normalised to the NLO $\mu=m_{t}$ result.
}                            
\end{center}
\vspace*{\hmysep}
\end{figure}
\begin{figure}
\begin{center}
\epsfig{file=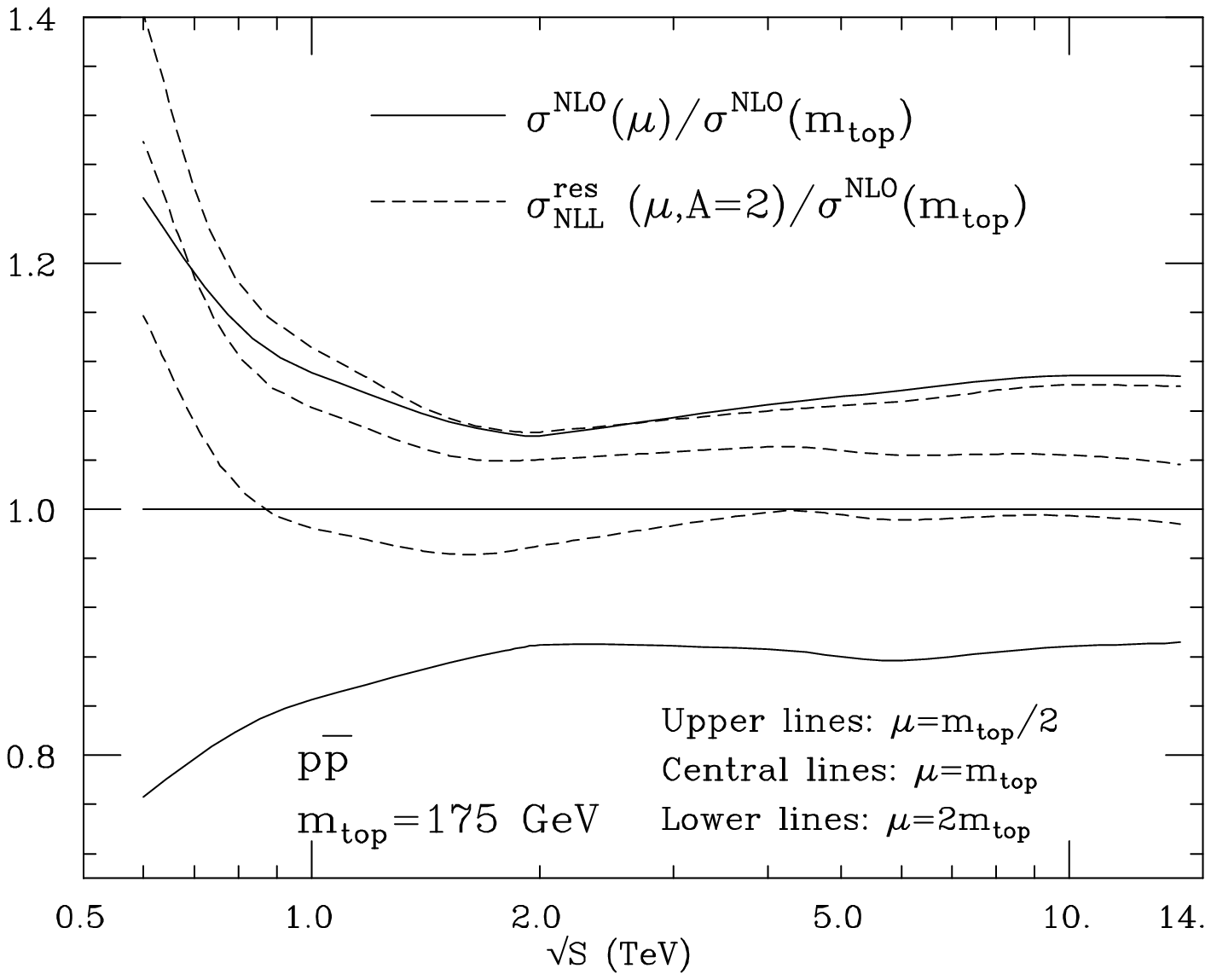,width=0.8\textwidth,clip=}
\vspace*{\mysep}
\ccaption{}{\label{fig:scaleA2} 
Same as fig.~\ref{fig:scaleA0}, but for $A=2$.}
\end{center}                                  
\end{figure}

\begin{figure}
\begin{center}
\epsfig{file=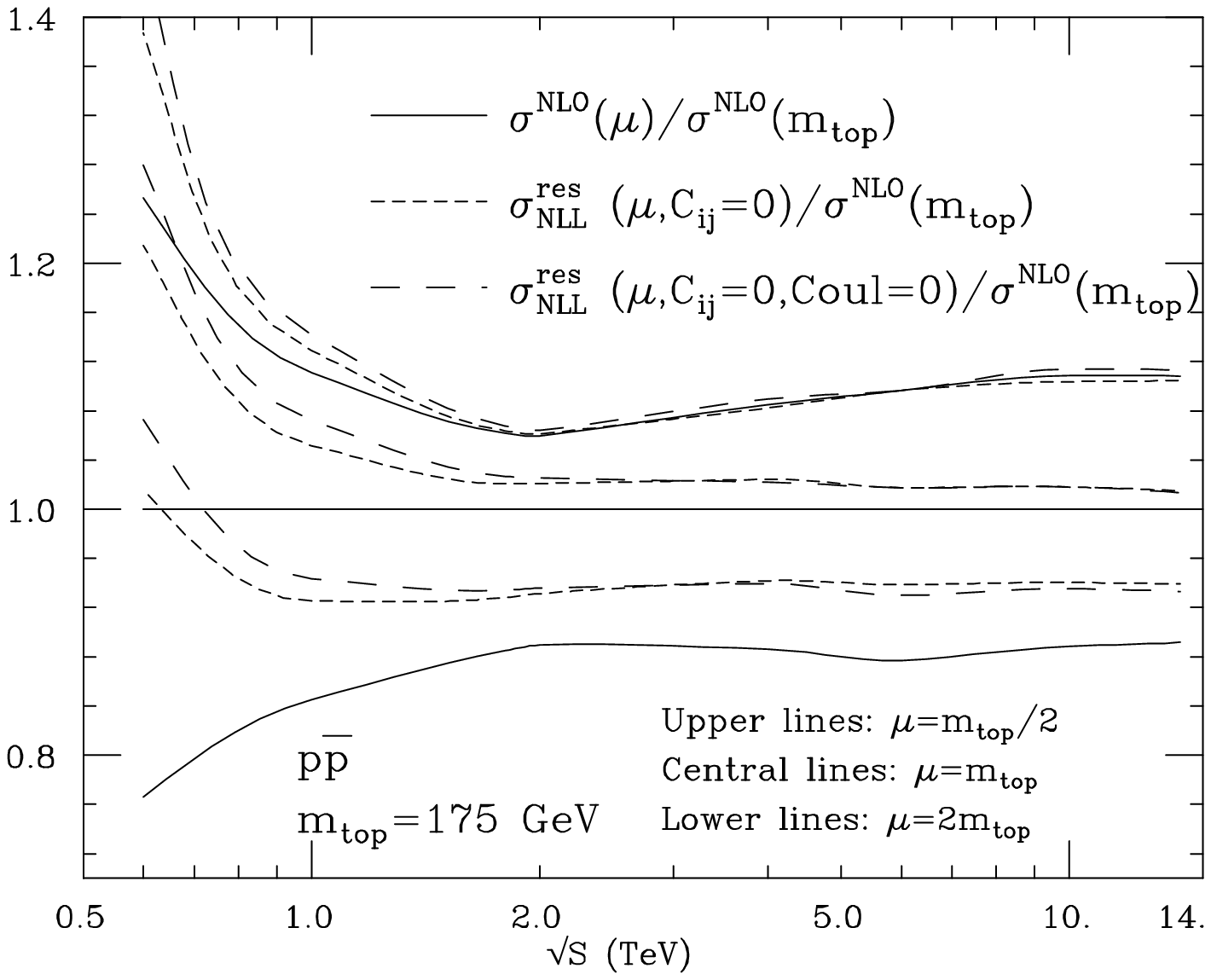,width=0.8\textwidth,clip=}
\vspace*{\mysep}
\ccaption{}{\label{fig:scaleC0} 
Same as fig.~\ref{fig:scaleA0}, but with the NNLL terms proportional to $C_{ij}$
set to zero (short-dashed lines). The dashed lines represent the same
calculation, with the Coulomb contributions in eq.~(\ref{fcorr}) removed.}
\end{center}                                  
\end{figure}

\begin{figure}
\begin{center}
\epsfig{file=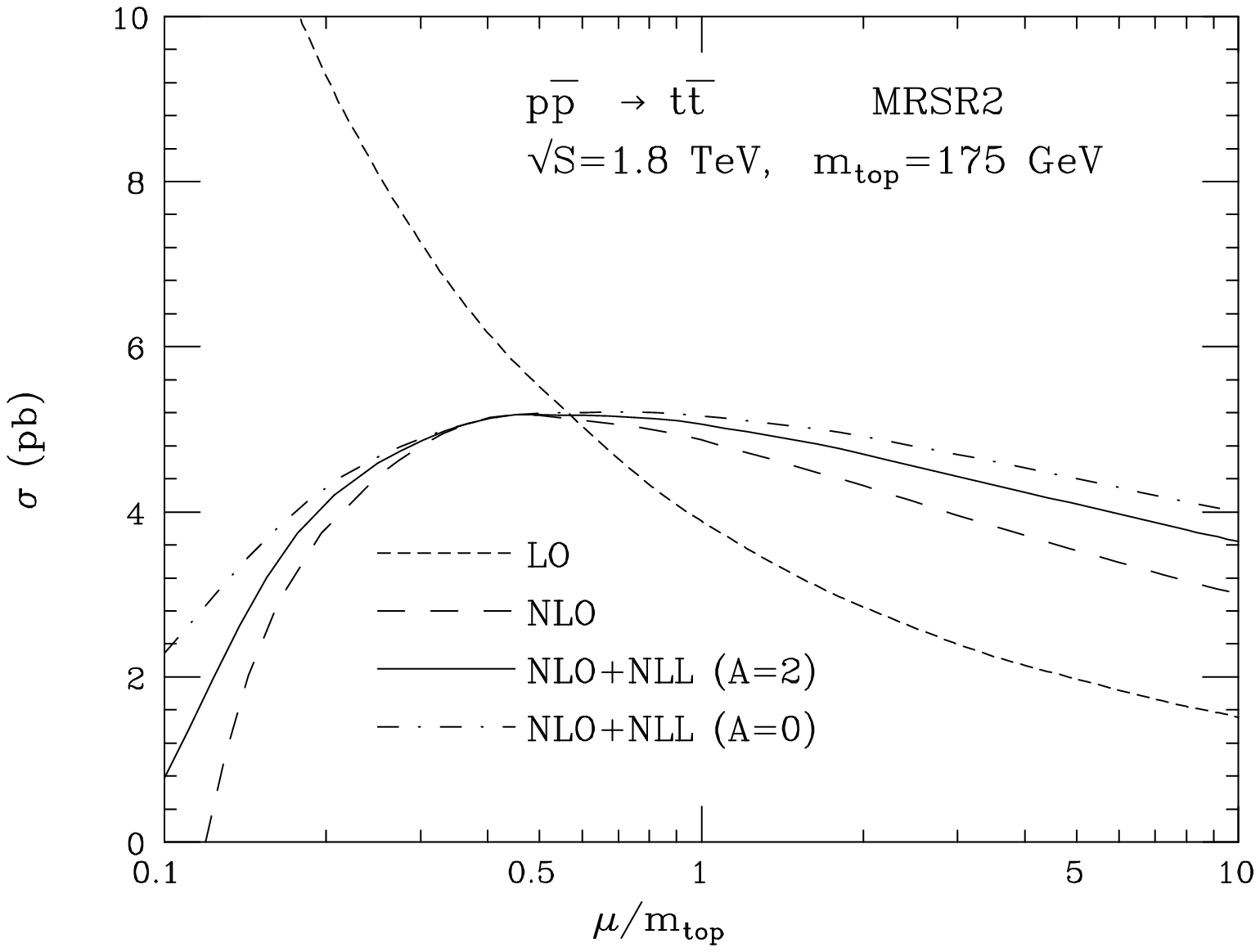,width=0.8\textwidth,clip=}
\vspace*{\mysep}
\ccaption{}{\label{fig:tscale} 
Scale dependence of the total $t\bar t$ production cross-section in $p \bar p$
collisions at
$\sqrt{S}=1.8$~TeV.
LO (short-dashed line), NLO (dashed line) and NLO+NLL (solid and dot-dashed
lines, for $A=2$ and $A=0$ respectively).}
\end{center}                            
\end{figure}
The scale dependence of the resummed cross-section, compared to the NLO one, is
given in figs.~\ref{fig:scaleA0} and \ref{fig:scaleA2}, corresponding to the
$A=0$ and $A=2$ cases, respectively.                    
Note the significant reduction in scale dependence, more marked in the $A=0$
case. More importantly, note that the band of variation of the resummed cross
section lies entirely within the band of variation of the NLO cross-section for
Tevatron energies and above. This shows that previous estimates of the
theoretical systematic uncertainty for the Tevatron cross-section were correct,
and can now be improved thanks to the NLL calculation presented here.

To display the importance of the inclusion of the NNLL $C_{ij}$ terms, we show
the same plot of the scale dependence with $C_{ij}=0$ in
fig.~\ref{fig:scaleC0}. While the scale sensitivity is slightly worse than in
the cases with $C_{ij}\ne 0$, there is still an important improvement over the
NLO behaviour. In the same figure, we also show the effect of neglecting the
contribution of Coulomb terms of order $\oafour$ and higher. With the exception
of the low-energy points, where Coulomb effects are very important because of
the closeness of the threshold, inclusion of the Coulomb terms in the
resummation is not a significant effect.
                                        
A more detailed representation of the scale sensitivity of the resummed 
top cross-section is shown in fig.~\ref{fig:tscale}, where we show the scale
dependence in the wide range $m_{t}/10 < \mu < 10 m_{t}$.

\begin{figure}
\begin{center}
\epsfig{file=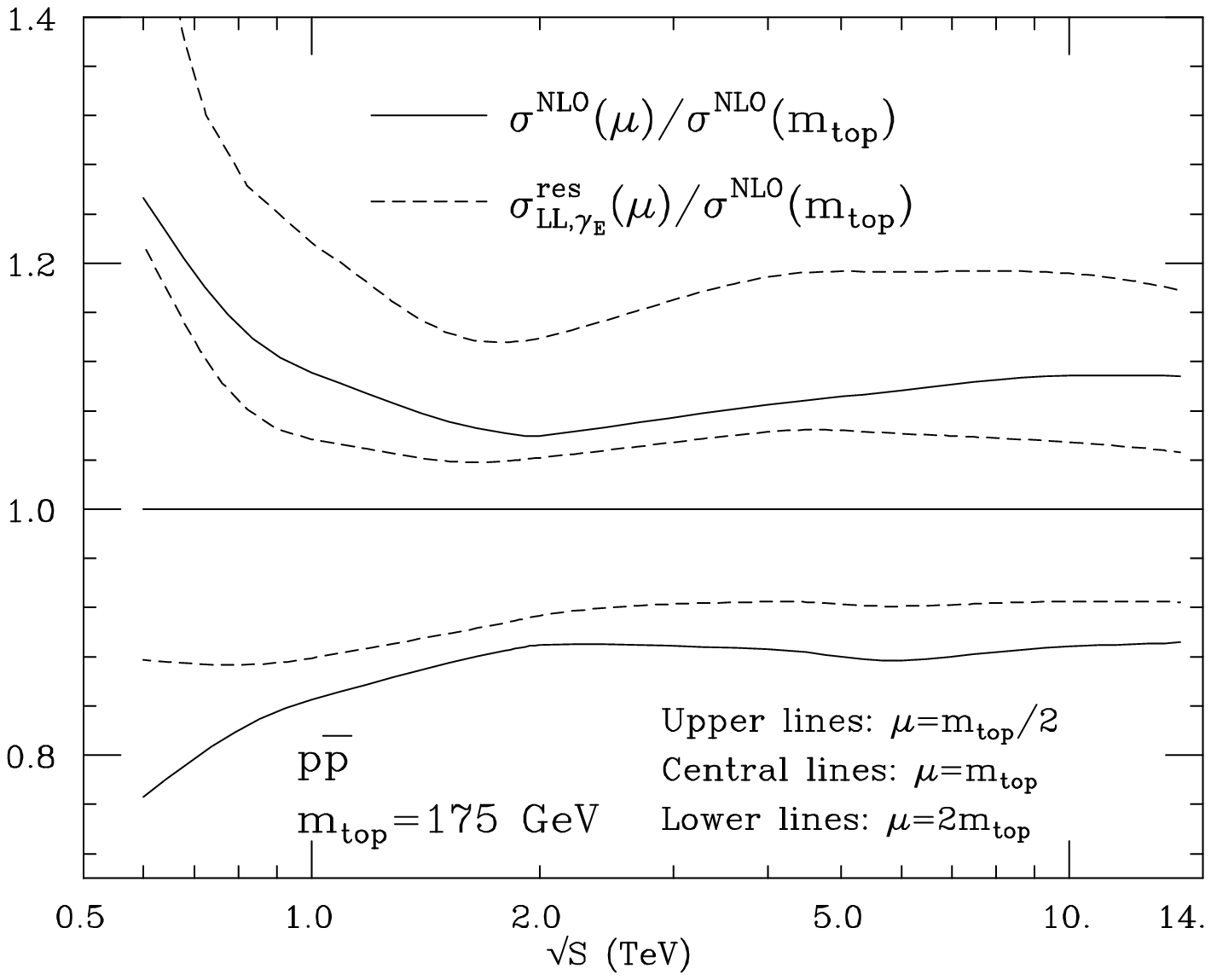,width=0.8\textwidth,clip=}
\vspace*{\mysep}
\ccaption{}{\label{fig:scaleA2eg} 
Same as fig.~\ref{fig:scaleA0}, but with the NLL result replaced with the
resummation of the leading logarithms only, and inclusion of the $\gamma_E$
terms defined by eq.~(\ref{eq:gammae}).}
\end{center}                  
\end{figure}

It was pointed out in ref.~\cite{CMNT2} that large NLL contributions to the
resummation of leading logarithms arise from the inclusion of NLL
corrections to the relation $1-z^N \sim \theta(1-z-1/N)$,
which
can be used to perform the integrals in eqs.~(\ref{dqq1})--(\ref{dgg8}). 
This relation,
which is valid at LL level, should be replaced at NLL order by
$1-z^N \sim \theta(1-z-e^{-\gamma_E}/N)$.
Neglecting terms beyond NLL accuracy, this replacement amounts to the
substitution\footnote{This is also equivalent to adding to 
$\ln\Delta^{LL}_{ij}$ the sole contributions proportional to $\gamma_E$ in the
expressions (\ref{g2fun}) for $g^{(2)}_{ij}$.}:
\beq    \label{eq:gammae}                   
     \Delta^{LL}_{ij,N} \to 
     \left( 1 \, + \, \gamma_E \, \frac{\partial}{\partial
\ln N} \right) \Delta^{LL}_{ij,N}\; .
\eeq                      
For the specific choice $\mu=\mt$, we showed in~\cite{CMNT2}
that the $\oacube$ truncation of the LL resummed calculation
provides a better agreement with the exact NLO
result for the partonic cross-section when this substitution is applied. 
This agreement is however accidental and limited to this
particular choice of the renormalization scale. 
Figure~\ref{fig:scaleA2eg} shows in fact that,
with the sole substitution in eq.~(\ref{eq:gammae}),
no improvement in the scale dependence is observed relative to the 
NLO calculation when the scale is varied.  Such an improvement can
only be obtained when the full set of NLL terms is included.
                                                            
\begin{figure}
\begin{center}
\epsfig{file=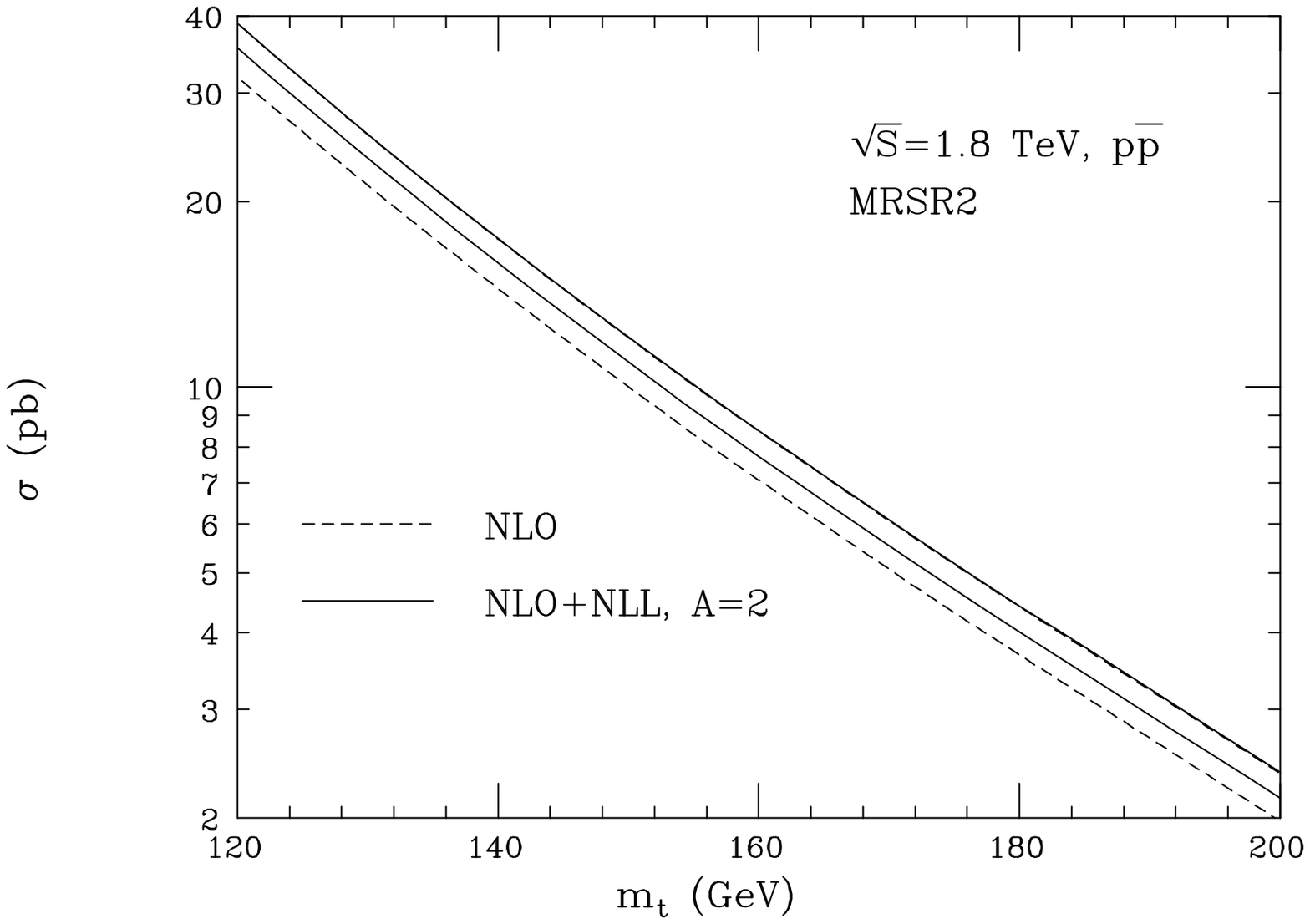,width=0.8\textwidth,clip=}
\vspace*{\mysep}
\ccaption{}{\label{fig:top18} 
Total $t\bar t$ production cross-section in $p\bar p$ collisions at
$\sqrt{S}=1.8$~TeV, as a function of the top-quark mass. Dashed lines: NLO
result; solid lines: NLO+NLL result. Upper lines: $\mu=m_{t}/2$;
lower lines: $\mu=2m_{t}$.}
\end{center}                 
\end{figure}

\begin{figure}
\begin{center}
\epsfig{file=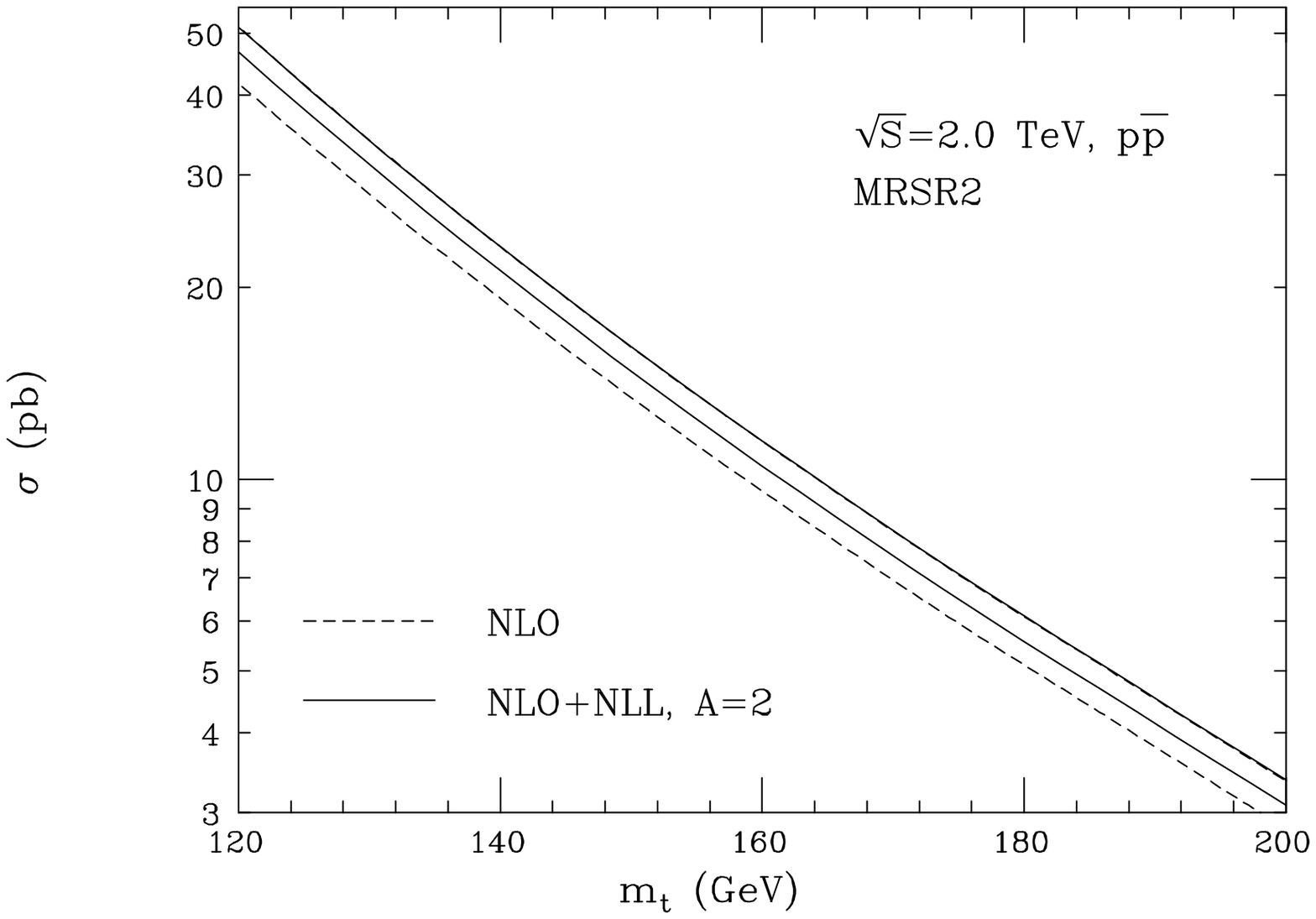,width=0.8\textwidth,clip=}
\vspace*{\mysep}
\ccaption{}{\label{fig:top20} 
Total $t\bar t$ production cross-section in $p\bar p$ collisions at
$\sqrt{S}=2.0$~TeV, as a function of the top-quark mass. Dashed lines: NLO
result; solid lines: NLO+NLL result. Upper lines: $\mu=m_{t}/2$;
lower lines: $\mu=2m_{t}$.
}                           
\end{center}                 
\end{figure}

\begin{figure}
\begin{center}
\epsfig{file=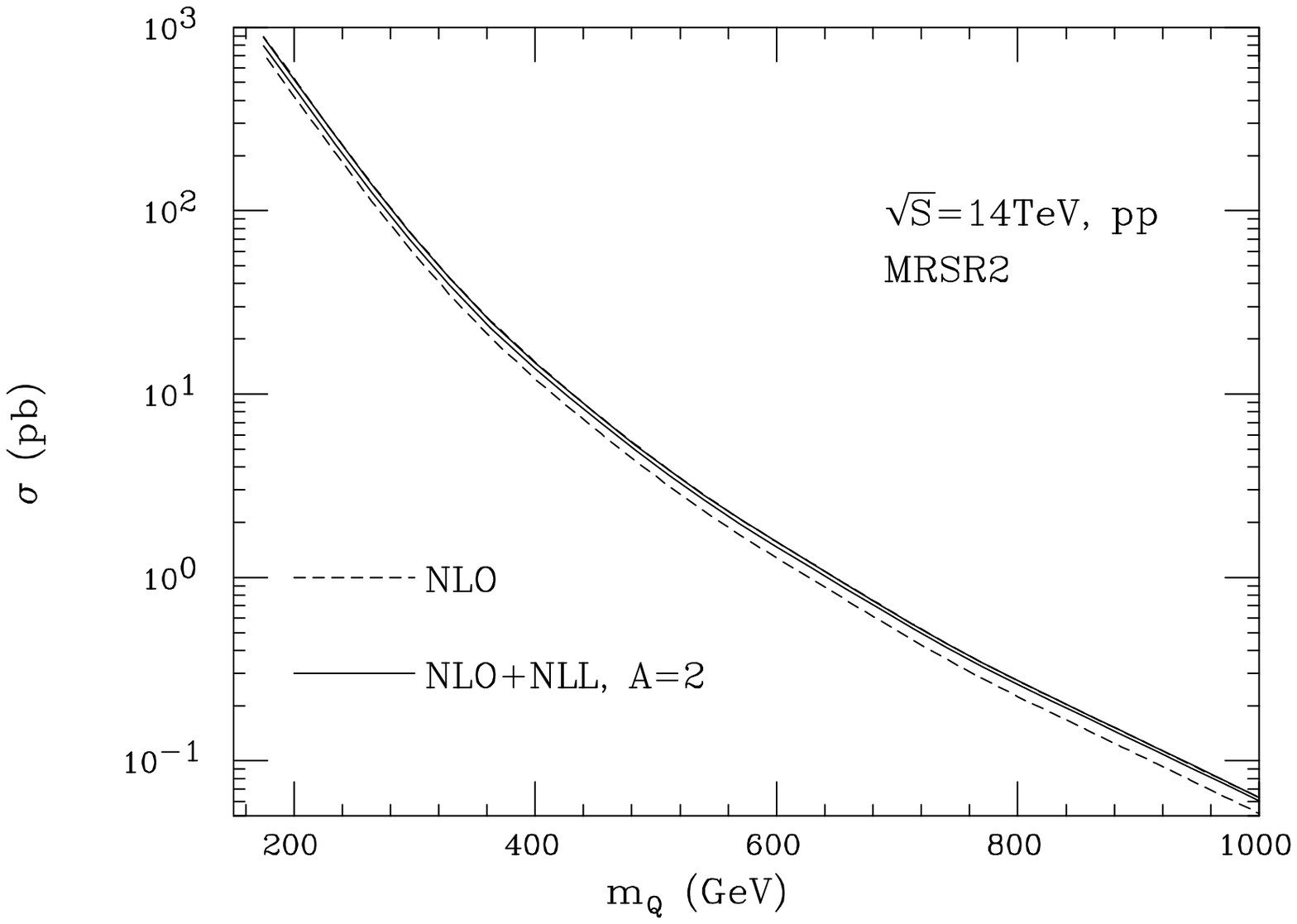,width=0.8\textwidth,clip=}
\vspace*{\mysep}
\ccaption{}{\label{fig:lhc} 
Total $Q\bar Q$ production cross-section in $p p$ collisions at
$\sqrt{S}=14$~TeV, as a function of the heavy-quark mass. Dashed lines: NLO
result; solid lines: NLO+NLL result. Upper lines: $\mu=m_{Q}/2$;
lower lines: $\mu=2m_{Q}$.
}                           
\end{center}                 
\end{figure}
            
The heavy-quark pair cross-section in $p\bar p$ collisions 
at $\sqrt{S}=1.8$ and 2~TeV are shown, as a
function of the heavy-quark mass, in figs.~\ref{fig:top18} and \ref{fig:top20}.
We chose the value $A=2$, which gives a more conservative estimate of the scale
uncertainty, and PDF set MRSR2 ($\as(M_{\rm Z})=0.119$). 
The two dashed lines
correspond to the NLO result, with 
$\mu=\mt/2$ and $2\mt$. The two solid lines correspond to the fully
resummed NLL result. At $\sqrt{S}=1.8$~TeV and for
$\mt=175$~GeV, we get a fully resummed result of
$\sigma_{t\bar t}=
5.06{+0.13 \atop -0.36}$~pb, compared to the fixed NLO one of
$\sigma_{t\bar t}=                                        
4.87{+0.30 \atop -0.56}$~pb (the central values representing the results with
$\mu=m_{t}$). The scale uncertainty of the resummed result is
reduced by almost a factor
of 2. 
The current experimental results from CDF~\cite{CDFtop}
and D0~\cite{D0top} are respectively:
$\sigma_{t\bar t}=
7.6{+1.8  \atop -1.5 }$~pb (CDF, at $\mt=175$~GeV), and
$\sigma_{t\bar t}=   
5.5\pm 1.8$~pb (D0, at $\mt=173.3$~GeV).
                                 
The NLO and fully resummed cross-sections for LHC ($pp$ collisions at
$\sqrt{S}=14$~TeV) are shown as a function of the heavy-quark mass in
fig.~\ref{fig:lhc}. 
{\renewcommand{\arraystretch}{1.8}
\begin{table}
\begin{center}
\begin{tabular}{|c|cc|cc|cc|} \hline           
& \multicolumn{2}{c|}{$p\bar p$ at $\sqrt{S}=1.8$~TeV }
& \multicolumn{2}{c|}{$p\bar p$ at $\sqrt{S}=2$~TeV }
& \multicolumn{2}{c|}{$p p$ at $\sqrt{S}=14$~TeV }\\
\cline{2-7}
 \mur=\muf   & NLO & NLO+NLL & NLO & NLO+NLL & NLO & NLO+NLL \\ \hline\hline
\mt/2 &   5.17  & 5.19   & 7.10  & 7.12 & 893  & 885  
\\ \hline                                             
\mt   &   4.87  & 5.06   & 6.70  & 6.97 & 803 & 833   
\\ \hline                                             
2\mt &   4.31  & 4.70   & 5.96  & 6.50 & 714 & 794
\\ \hline                                          
\end{tabular}
\ccaption{}{\label{tab:topxs} Total $t\bar t$ cross-sections ($m_t=175$~GeV)
at the Tevatron and LHC, in pb. PDF set MRSR2.} 
\end{center}      
\end{table} }

{\renewcommand{\arraystretch}{1.8}
\begin{table}
\begin{center}
\begin{tabular}{|c|c|c|c|c|c|c|} \hline
& $\alpha_s^2$ & $\alpha_s^3$ & $\alpha_s^4$ & $\alpha_s^5$ &
$\alpha_s^6$ & $\alpha_s^{\ge 7}$  \\ \hline \hline
$q \bar q$ &   3590 & 766  & 60.2 & 2.4 & $-0.6$ & $-0.08$ \\
$gg$       &    298 & 264  & 98.1 & 26.1 & 5.8 & 1.5 \\
$q\bar q+gg$ & 3888 &1030 & 158  & 28.5 & 5.1 & 1.4 \\ \hline
\end{tabular}                                      
\ccaption{}{\label{tab:toppert} Contributions to the 
total $t\bar t$ cross-sections  (in fb) at the Tevatron (1.8~TeV)
from higher orders in the expansion of
the NLL resummed result, with
$\mu=m_t$, $m_t=175$ GeV and PDF set MRSR2.    
The second column gives the {\em exact} NLO result.
The last row only includes the sum of the $q \bar q+gg$ channels.}
\end{center}      
\end{table} }
In Table~\ref{tab:topxs} we collect the values of the
cross-sections for $\mt=175$~GeV at the Tevatron and LHC.
The rapid convergence of the higher-order corrections is displayed in
table~\ref{tab:toppert}, for the case of top production at the Tevatron.
The sum of all entries in each row corresponds to the minimal prescription,
while each fixed order term does not have any ambiguity due to the choice
of the contour for the Mellin transformation in eq.~(\ref{MPHQ}).
This supports the validity of the minimal prescription,
since the truncated resummed expansion converges to it very rapidly.

{\renewcommand{\arraystretch}{1.8}
\begin{table}
\begin{center}
\begin{tabular}{|c|cc|cc|cc|} \hline
& \multicolumn{2}{c|}{$m_b$=4.5 GeV} & \multicolumn{2}{c|}{$m_b$=4.75 GeV} 
& \multicolumn{2}{c|}{$m_b$=5.0 GeV} \\                             
\cline{2-7}                   
 \mur=\muf   & NLO & NLO+NLL & NLO & NLO+NLL & NLO & NLO+NLL \\ \hline\hline
\mb/2 &   27.4 & 27.3 & 17.6  & 17.7 & 11.2  & 11.5 \\     
\hline                                             
\mb   &   15.7  & 22.8 & 9.90  & 14.9 & 6.26  & 9.75 \\
\hline                                              
2\mb &    8.74  & 18.0 & 5.47  & 11.7 & 3.43  & 7.63 \\
\hline                                              
\end{tabular}
\ccaption{}{\label{tab:herab1} Total $b\bar b$ cross-sections (in nb)
at HERAB ($pp$ at $\sqrt{S}=39.2$~GeV), as a function of the the $b$ mass 
$m_b$. PDF set MRSR1.}
\end{center}      
\end{table} }

{\renewcommand{\arraystretch}{1.8}
\begin{table}
\begin{center}
\begin{tabular}{|c|cc|cc|cc|} \hline
& \multicolumn{2}{c|}{$m_b$=4.5 GeV} & \multicolumn{2}{c|}{$m_b$=4.75 GeV} 
& \multicolumn{2}{c|}{$m_b$=5.0 GeV} \\                             
\cline{2-7}
 \mur=\muf   & NLO & NLO+NLL & NLO & NLO+NLL & NLO & NLO+NLL \\ \hline\hline
\mb/2 &   45.2  & 44.9 & 28.6  & 28.9 & 18.0  & 18.6 \\     
\hline                                              
\mb   &   22.2  & 35.3 & 13.9  & 22.9 & 8.71  & 14.9 \\
\hline                                              
2\mb &    11.2  & 26.1 & 6.94  & 17.0 & 4.31  & 11.0 \\
\hline                                              
\end{tabular}
\ccaption{}{\label{tab:herab2} Total $b\bar b$ cross-sections  (in nb)
at HERAB ($pp$ at $\sqrt{S}=39.2$~GeV), as a function of the the $b$ mass 
$m_b$. PDF set MRSR2.}
\end{center}      
\end{table} }

Potentially large resummation effects should be expected in $b\bar b$
production at fixed target energies, due to the closeness of the threshold. As
an example, we collect in      
Table~\ref{tab:herab1} the values of the $b\bar
b$ total cross-section calculated using the MRSR1 set ($\as(M_{\rm Z})=0.112$)
for $pp$ collisions at $\sqrt{S}=39.2$~GeV,
the configuration of the upcoming HERAB experiment at DESY. We show both the
NLO results and the fully resummed ones, for several values of the
bottom quark mass and for the three standard choices of renormalization scale.
The resummation corrections are large and positive for large values of the
renormalization scale, but become small -- and eventually negative --
for the smaller values of the
renormalization scale. As a result, the overall scale dependence of the
cross-section is significantly reduced 
in the fully resummed calculation, and
the overall uncertainty band is fully included in the uncertainty band of the 
NLO calculation.
\begin{figure}
\begin{center}
\epsfig{file=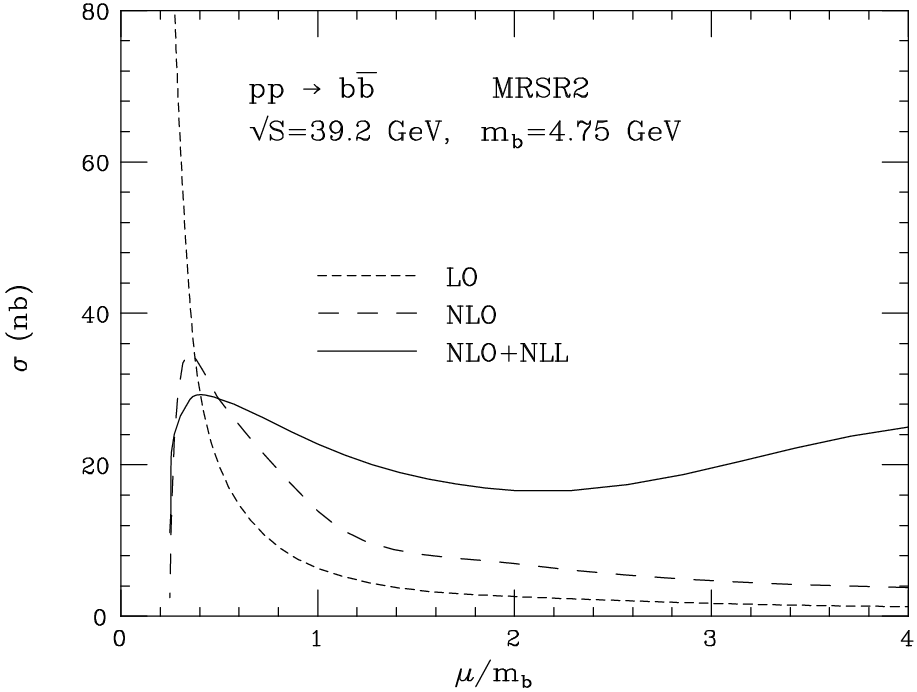,width=0.8\textwidth,clip=}
\vspace*{\mysep}
\ccaption{}{\label{fig:bscale} 
Scale dependence of the total $b\bar b$ production cross-section at HERAB, 
at LO (dotted line), NLO (dashed line) and NLO+NLL (solid line).}
\end{center}                 
\end{figure}

The NLO cross-section, and the resummation effects, are largely enhanced by the
choice of the MRSR2 set, for which the value of the coupling constant is
larger. The results are shown in Table~\ref{tab:herab2}. The conclusions
regarding the improved stability of the resummed result, as well as the 
reliability of the uncertainty estimate obtained from the scale variation of
the NLO calculation, are nevertheless unchanged. 
A plot of the scale dependence of the LO, NLO and resummed cross-sections, for
$m_b=4.75$~GeV and PDF set MRSR2 is shown in fig.~\ref{fig:bscale}.
                                                                   
As a comparison, we give here the value of the bottom-pair cross-section at
the Tevatron, where we expect much smaller resummation effects.
The NLO cross-section 
at the Tevatron ($\mb=4.75$~GeV, MRSR2) is:       
$\sigma=74.8,\, 56.2, \, 46.6$~$\mu$b for $\mu=\mb/2,\, \mb, \, 2\mb$.
Including resummation we obtain:
$\sigma=74.2,\, 57.6, \, 52.2$~$\mu$b for $\mu=\mb/2,\, \mb, \, 2\mb$. Some
reduction in the scale dependence can be seen, but not at the level exhibited
by the HERAB results. This is consistent with the fact that $b$ production at
the Tevatron takes place far away form the threshold, and the 
class of logarithms resummed in this work cannot be expected to give a dominant
contribution.

\begin{figure}
\begin{center}
\epsfig{file=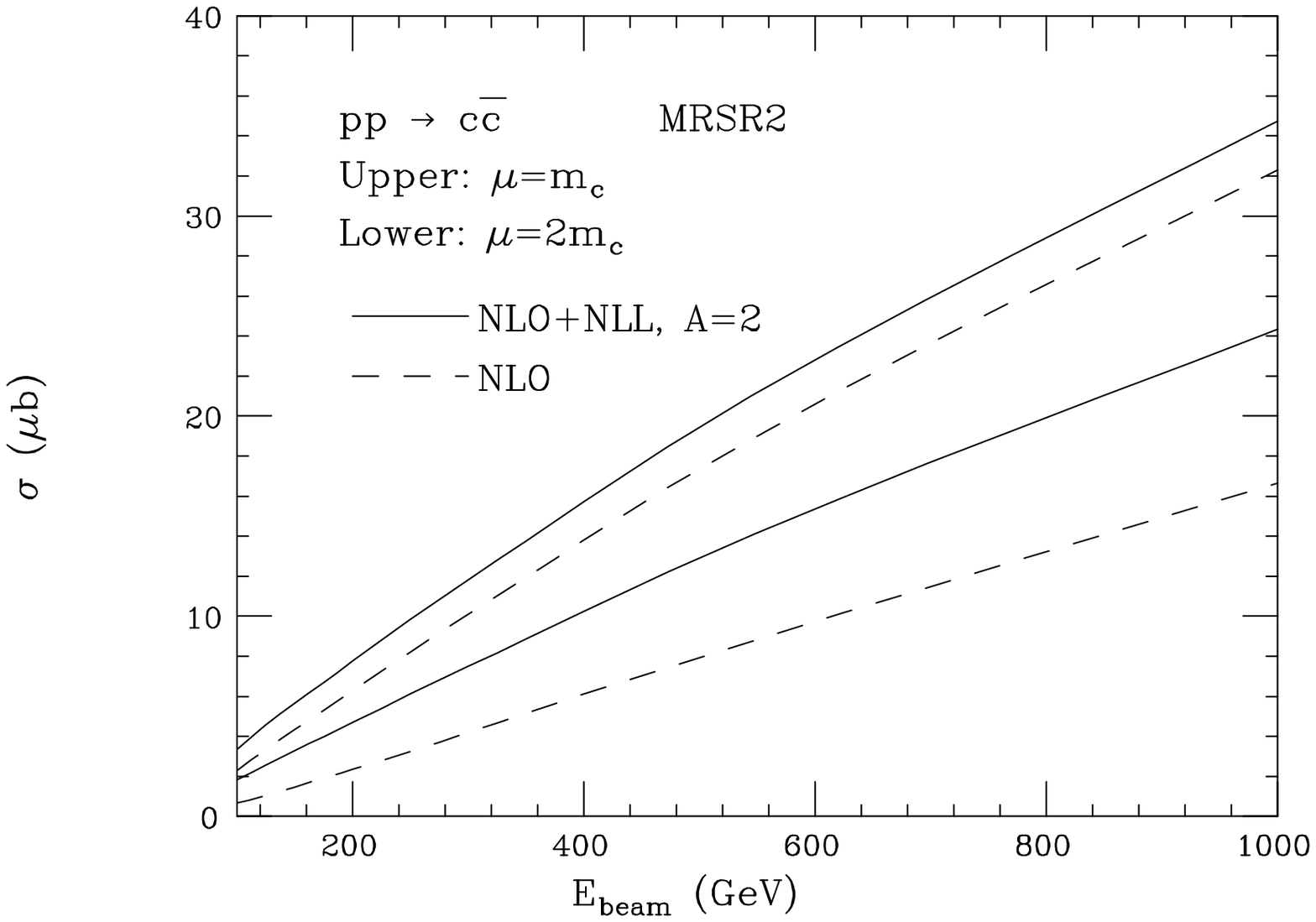,width=0.8\textwidth,clip=}
\vspace*{\mysep}
\ccaption{}{\label{fig:charm} 
Total $c\bar c$ cross-section in fixed-target $pp$ collisions, as a function of
beam energy, at the NLO and NLO+NLL.}                           
\end{center}                           
\end{figure}
                           
To conclude, we show in fig.~\ref{fig:charm} the effect of resummation on the
total cross-section for fixed-target hadroproduction of charm-quark pairs.
The typical values of $4 m_c^2/S$ at fixed-target energies (where $\sqrt{S}$ is
in the range 20--40~GeV) are close to the values for top-pair production at the
Tevatron. Contrary to the top case, however, charm production is dominated in
$pp$ collisions by the $gg$ annihilation channel, due to the smallness of the
antiquark sea. Considering in addition the larger size of the strong coupling
constant at scales of the order of the charm mass, we therefore
expect the resummation corrections to be large.
Results are plotted as a function of the beam energy, for proton-induced
reactions. The interval for the scale variation is limited to the range $m_c
< \mu < 2 m_c$, since the parton densities we use (MRSR2) are limited to the
domain $\mu^2>1.25$~GeV$^2$.  Once
more, the resummed result shows a significant reduction in the scale 
dependence over the NLO result.
                                                                    
\section{Discussion and conclusions}                                     
We presented in this paper the first calculation of the heavy-quark
hadro-production total cross-section accurate up to, and including,
next-to-leading threshold-enhanced logarithms, resummed at all orders of
perturbation theory. One can show that our analytical results 
for the resummed total cross-section are consistent with those obtained in
in~\cite{kidon} for the cross-section at fixed invariant mass of the
heavy-quark pair. When the resummed partonic 
cross-sections are convoluted with hadronic parton densities, a significant
improvement is observed in the stability of the results with respect to 
changes of the renormalization scale. In the case of top-quark production at
the Tevatron, the corrections relative to the NLO calculation are of the
order of 9\% for $\mu=2m_{t}$, 4\% for $\mu=m_{t}$, and 0.5\% for 
$\mu=m_{t}/2$.
The result for $\mu=m_{t}$ can be summarized as follows.
Including LL terms we get a 0.5\% correction; the correction increases
to 4\% when higher-order                                              
terms proportional to $\gamma_E$ are included; 
when all NLL contributions are included      
the correction decreases to 2.6\%; when the Coulomb term
is added the correction decreases to 2.1\%; when the first NNLL terms
(the $C_{ij}$ constants) are 
included the correction grows to 4\% for the choice
$A=2$, to 6\% for the choice $A=0$. As we have seen,
the $A=0$ choice is somewhat disfavoured, since it
overestimates the $\oacube$ cross-section above threshold.
We emphasize that the overall compression of the uncertainty band
toward the high cross-section values is significant and always present
once we include the NLL terms, regardless of 
the presence of the $C_{ij}$ and Coulomb terms, and regardless
of our choice of the value of $A$.            
                 
In the case of top productio at the Tevatron,
it was shown in ref.~\cite{CMNT2} that
inclusion of the sole NLL terms proportional to $\gamma_E$ provided an
improvement in the agreement  with the $\oacube$ partonic cross-section for the
choice of scale $\mu=m_{t}$. This
improvement is however accidental, and indeed, as shown in this paper, it is
not sufficient to improve the accuracy of the resummed calculation when the
renormalization scale is changed to different values. Only at the full NLL
level a significant reduction in the scale dependence is found.
Similar results are obtained at higher energies, including the energy of the
future LHC collider. 

The improvement in the predictive power allowed by the calculation we presented
in this work is even more impressive when one considers the case of the 
bottom-quark production near threshold, at the future HERAB experiment. In this
case the size of the strong coupling constant gives rise to very large
perturbative corrections, which result in a factor of 4 uncertainty in the
NLO result at a fixed value of the bottom mass and parton
densities. Inclusion of resummation effects stabilizes the predicted
cross-section at the level of $\pm 50$\%.
As in the case of the top cross-sections, the evaluation of the fully resummed
cross-sections shows that the estimates of the theoretical
uncertainty at the NLO obtained by varying the renormalization scales in
the range $m_Q/2 < \mu < 2 m_Q$ were correct. While the resummed result has a
much smaller uncertainty than the fixed-order one, the resulting range of
predictions is always included in the uncertainty band estimated at the NLO.

\noindent {\bf Acknowledgement:} We thank Matteo Cacciari for pointing
out an error in Fig.~14 of the first version of this paper.

\end{document}